\documentclass[
  journal=largetwo,
  manuscript=article-type,
  year=2023,
  volume=,
]{cup-journal}

\usepackage{amsmath}
\usepackage{aas-macros}
\usepackage[nopatch]{microtype}
\usepackage{booktabs}

\title{Inviscid protostellar disc  ring formation and high density ring edges due to the ejection and subsequent infall of material onto a protostellar disc.}

\author{K. Liffman}
\affiliation{Centre for Astrophysics and Super Computing, Swinburne University of Technology, Hawthorn, VIC 3122, Australia}
\email[K. Liffman]{kliffman@swin.edu.au}


\begin{document}

\begin{abstract}
Discs of gas and dust are ubiquitous around protostars. Hypothetical viscous interactions within the disc are thought to cause the gas and dust to accrete onto the star. Turbulence within the disc is theorised to be the source of this disc viscosity. However, observed protostellar disc turbulence often appears to be small and not always conducive to disc accretion. In addition, theories for disc and planet evolution have difficulty in explaining the observed disc rings/gaps which form much earlier than expected.

Protostellar accretion discs are observed to contain significant quantities of dust and pebbles. Observations also show that some of this material is ejected from near the protostar, where it travels to the outer regions of the disc. Such solid infalling material has a relatively small amount of angular momentum  compared to the material in the disc. This infalling material lowers the angular momentum of the disc and should drive a radial flow towards the protostar.

We show that the local radial accretion speed of the disc is proportional to the mass rate of infalling material onto the disc. Higher rates of infall onto the disc implies higher radial accretion disc speeds. As such, regions with high rates of infall of gas, dust, and pebbles onto the disc will produce gaps on relatively short timescales in the disc, while regions associated with relative low rates of infalling material will produce disc rings. As such, the inner edge of a disc gap will tend to have a higher surface density, which may enhance the probability of planet formation. In addition, the outer edge of a disc gap will act as a dust trap and may also become a site for planet formation.

For the early Solar System, such a process may have collected O$^{16}$-poor forsterite dust from the inner regions of the protosolar disc and O$^{16}$-rich CAIs and AOAs from the inner edge regions of the protosolar disc, thereby constructing a region favourable to the formation of prechondritic planetesimals.

\bf{Accepted for publication in PASA}
\end{abstract}

\section{Introduction }
\label{sec:int}

When first formed, protostellar systems consist of a central protostar surrounded by a disc of gas and dust, where the disc accretes material onto the star \citep{2011ARA&A..49...67W}. The source of the energy loss that allows material to flow from the disc to the star has long been a source of theoretical speculation
\citep{2019NewA...70....7M}. Most models use magnetic or hydrodynamic instabilities \citep{1991ApJ...376..214B,2019EAS....82..391F} to create turbulence which, in turn,  produces viscous effects within the disc \citep{2002apa..book.....F}. The standard $\alpha$ disc model being based on various assumptions regarding disc turbulence \citep{1973A&A....24..337S}. 

These are useful models that are probably applicable to protostellar discs in certain regions and at particular stages of formation, but, to date, there is little observational evidence to support the presence of wide scale, significant turbulence in many protostellar discs \citep{2017ApJ...843..150F,2018ApJ...864..133T}. 

In the absence of wide scale disc turbulence, disc magnetic fields have long been suggested as a plausible driver of disc accretion, \citep{2016ApJ...821...80B,2021SciA....7.5967W}. Unfortunately, it is not easy to detect magnetic fields in protostellar systems, however, meteorites may provide additional information regarding magnetic fields in the solar protostellar system \citep{2021SciA....7.6928B}.

Observations suggest that sub-mm and mm to cm-sized particles are abundant in protostellar discs \citep{2019ApJ...886..103O}. These observations are also consistent with the observed high abundance of such particles (e.g., chondrules, Calcium Aluminium Inclusions (CAIs), Amoeboid Olivine Aggregates (AOAs) \&c) found in primitive meteorites \citep{2017ASSL..445..161B}. 

Standard disc theory, would suggest that all these particles should quickly drift into the protostar \citep{1977MNRAS.180...57W}. It is possible that turbulence within discs can transport particulate material from the inner to the outer regions of the disc \citep{2022ApJ...940..117Z}. However, the observed low disc turbulence and prominent gaps in even the youngest protostellar discs \citep{2019ApJ...872..112V} suggests that this upstream turbulent advection of particulate material within protostellar discs may be of limited effect.

In addition, Spitzer Space Telescope and James Webb Space Telescope observations also demonstrate that crystalline, micron-sized particles are ejected from the inner regions of protostars and travel radially across the face of the disc and reenter the disc some au from the protostar or are completely ejected from the protostellar system \citep{2011ApJ...733L..32P,2012ApJ...744..118J,2019ApJ...887..156A,2023ApJ...945L...7K}.  These particles from the inner regions of the disc will tend to have little or no angular momentum relative to material in the outer regions of the disc. When these ejected particles re-enter the disc at greater distances, then the angular momentum of the outer disc is reduced, and the outer disc regions will move radially inwards towards the protostar.

\begin{figure}
\centering
\includegraphics[width=\textwidth]{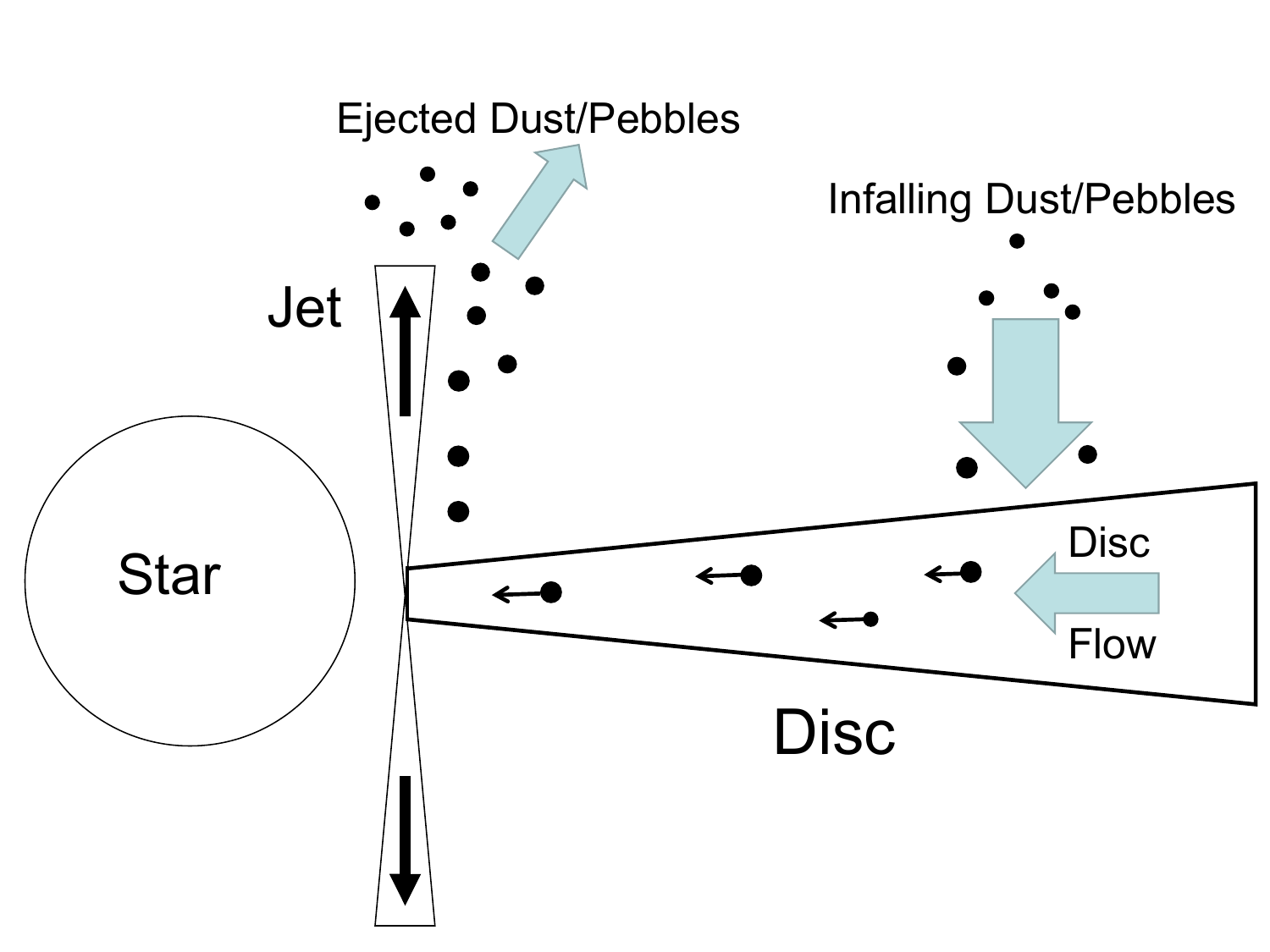}
\caption{Low angular momentum particles are ejected from the inner disc and land in the outer regions of the disc, where they are returned to the inner disc by radial inflow and/or radial drift, thereby renewing the cycle. The removal of angular momentum from the outer disc produces disc flow towards the star. For the disc as a whole, this is likely to be a small effect and more conventional sources of disc accretion are probably required to drive accretion onto the protostar.}
\label{fig:Virtuous_cycle}
\end{figure}

It is plausible that the entrained dust and small pebbles can move with the gas or drift back to the inner disc region where the outflow is located. These particles may then be re-ejected to the outer regions of the disc and a repeating, virtuous cycle is produced that drives disc accretion into the inner regions of the accretion disc, whereupon conventional sources of viscosity can drive accretion onto the protostar (Figure \ref{fig:Virtuous_cycle}). 

This cycle can potentially replenish the pebbles in the disc \citep{2014MNRAS.440.3545H} and thereby counter the expected relatively fast drift of pebbles into the protostar \citep{1977MNRAS.180...57W,2007A&A...469.1169B,2008A&A...480..859B}. However, the deduced mass accretion rates from this effect will likely be small relative to what is observed. This occurs because the mass of dust/pebbles in the disc will be of order 1\% of the disc mass. This implies that the solid material has to be cycled around approximately 100 times to have a major influence in driving material onto the protostar.

However, this infalling material is observed, at least in one case, to be heterogeneously distributed across the accretion disc \citep{2012ApJ...744..118J,2019ApJ...887..156A,2023ApJ...945L...7K}. Some theoretical models (e.g., \citet{2019ApJ...882...33G}) also explicitly predict that disc winds will deposit particulate material outside the outflow region. So, there may be discrete regions of the disc that are strongly affected by infalling material.

\begin{figure}
\centering
\includegraphics[width=\textwidth]{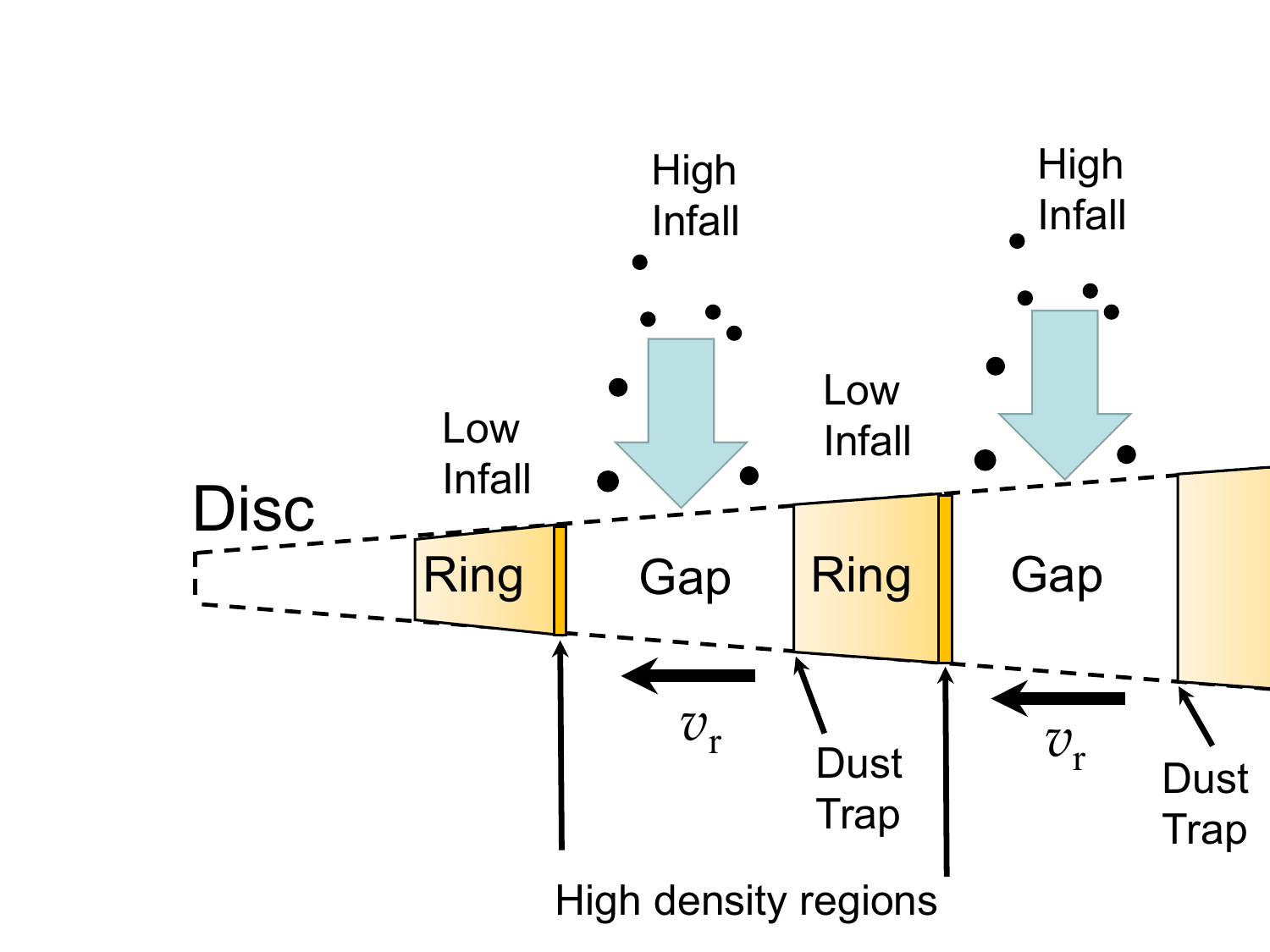}
\caption{Sections of the disc that suffer high rates of infalling material will have a higher radial disc speed, $v_{\rm r}$, which will help produce disc gaps on a relatively short timescale. Conversely, low infall rates will produce low radial disc speeds and rings of disc material. The material in the gap regions will flow into the ring regions. The outer regions of the rings will, therefore, have enhanced densities and pressures, while the inner regions of the rings will act as dust traps, so both ring edges may provide fertile ground for planet formation.}
\label{fig:Rings_and_gaps}
\end{figure}

The theory presented here shows that the radial speed of disc material is proportional to the local mass rate of infall. If there are spatial or temporal inhomogeneities in this infalling material, then there will also be variations in the local radial speed of the disc moving towards the protostar. Higher mass infall regions will have a higher local radial disc speed, while lower mass infall regions will have a relatively lower radial disc speed. Such inhomogeneities will produce gaps and rings in the disc on a relatively short timescale. Higher mass infall will produce gaps, while lower mass infall will produce rings. Due  to this accretion behaviour, the material in the gaps will flow into the outer edges of the rings and these regions will have enhanced densities and pressures, which may aid in planet formation (Figure \ref{fig:Rings_and_gaps}). In addition, the outer edges of the gaps will tend to have a pressure profile which will trap dust \citep{2012A&A...538A.114P}. So, planetesimal formation may become more likely at both the inner and outer edges of the gaps.

In this initial model, we do not consider the back reaction of the inward drifting pebbles on the disc gas (e.g., \citet{1986Icar...67..375N}). This back reaction on the gas would tend to cause the entire gas disc to increase in size and help conserve angular momentum for the entire disc. As such, we do not consider the angular momentum behaviour of a whole disc, but only consider the affect of angular momentum loss in discrete sections of a disc.

We also do not consider infall of material from a molecular cloud core onto a nascent protostellar disc. From \citet{2022ApJ...928...92K}, it is clear that such infalling material can intercept the disc in regions with significantly different angular momentum. \citet{2022ApJ...928...92K} use a full hydrodynamic code (albeit with approximations in the boundary conditions and code to make the problem computationally tractable) to illustrate how discrete, low angular momentum infalling material from a molecular cloud core can produce gaps and rings in a protostellar disc. The simplified disc evolution equations derived in this paper are also applicable as they apply to any case where material rains down upon a disc. 

This paper is organised as follows: in \S\ref{sec:disceqn} we derive the infall disc equations; in \S \ref{sec:Ring/Gaps} these equations are solved analytically to illustrate ring and gap formation on relatively short timescales in a protostellar disc; in \S \ref{sec:discussion} we discuss the ''chicken and egg'' problem of ring/gap versus planet formation, where we suggest that formation of rings and gaps is the necessary foundation for planet formation.

\section{Disc Equations with Infall}
\label{sec:disceqn}

As discussed in \citet{1997ApJ...479..740F} and \ref{sec:infall_eqns}, the equation for mass conservation in a disc with infalling material is:
\begin{equation}
   r \frac{\partial \Sigma}{\partial t} = -\frac{\partial \left(r v_{\rm r} \Sigma \right)}{\partial r} + r\dot{\Sigma}_{\rm i} \ ,
    \label{eq:mass_con}
\end{equation}

with $r$ the radial cylindrical coordinate, $t$ the time, $\Sigma(r,t)$ the surface density of the disc, $v_{\rm r}$ the radial velocity of the flow of disc material and $\dot{\Sigma}_{\rm i}(r,t)$ the total rate of infall of material per unit area and time onto the disc, where we note that, in this paper, the infall is assumed to be symmetric on {\it both} sides of the disc.

The infalling material may arise from the ambient medium around the protostellar system. The derived equations do not discriminate between gaseous and particulate material. However, in this paper, we assume that it arises due to the action of the protostellar jet flow and/or disc winds that eject solid material to different heights above the disc and different distances from the protostar. As such, the particulate material will return to the disc on a range of timescales. It is also possible that solid material will be ejected from the disc via disc winds and/or a semiquiescent jet flow. Whatever the exact ejection mechanism, we note that observations require particulate material to be ejected from the inner disc to the outer regions of the disc.

The angular momentum equation is tentatively:
\begin{equation}
   r \frac{\partial \left( \Sigma r^2 \Omega \right) }{\partial t} = -\frac{\partial \left( v_{\rm r}  \Sigma r^3 \Omega \right)}{\partial r} + \frac{1}{2\pi}\frac{\partial Q}{\partial r}+ \dot{\Sigma}_{\rm i} r^3 \Omega_{\rm i} \ ,
    \label{eq:ang_con1}
\end{equation}
where $\Omega(r,t)$ and $\Omega_{\rm i}(r,t)$ are the angular speeds of the disc and infalling material, respectively. $Q(r,t) = 2\pi \nu \Sigma r^3 \frac{\partial \Omega}{\partial r}$ is the disc viscous torque, with $\nu$ the disc viscosity.

We suppose that the standard astrophysical disc viscosity is only significant and applicable at the initial stage(s) of disc formation or in the very inner regions of the disc. it is certainly possible that early forming planets or companion stars can also induce disc accretion onto the protostar, but we shall ignore such possibilities. Given these constraints, disc turbulence is taken to be negligible and $\nu \approx 0$, which implies that $Q \approx 0$ for most protostellar discs. The disc angular momentum equation becomes:

\begin{equation}
   r \frac{\partial \left( \Sigma r^2 \Omega \right) }{\partial t} = -\frac{\partial \left( v_{\rm r}  \Sigma r^3 \Omega \right)}{\partial r} +  \dot{\Sigma}_{\rm i} r^3 \Omega_{\rm i} \ ,
    \label{eq:ang_con2}
\end{equation}

As derived in \ref{sec:disc_infall}, the disc radial speed, $v_{\rm r}$, from equations (\ref{eq:mass_con}) and (\ref{eq:ang_con2}), is:
\begin{equation}
    v_{\rm r} = \frac{1}{\frac{\partial (r^2 \Omega)}{\partial r}}\frac{\dot{\Sigma}_{\rm i}}{\Sigma}r^2(\Omega_{\rm i} - \Omega) \ .
\label{eq:rad_drift1}
\end{equation}
For Keplerian angular disc speed, (i.e., $\Omega \approx \sqrt{{\rm G}M_*/r^3}$, where $G$ is the gravitational constant and $M_*$ the mass of the central object) we have:

    \begin{equation}
      \begin{aligned}
    v_{\rm r} &= 2r\frac{\dot{\Sigma}_{\rm i}}{\Sigma}\left(\frac{\Omega_{\rm i}}{\Omega} - 1\right) \\
    &\approx 0.95 \ \text{mm s}^{-1}\left(\frac{r}{1 \text{au}} \right) \left(\frac{\dot{\Sigma}_{\rm i}/\Sigma}{10^{-7}/\text{yr}} \right)\left(\frac{\Omega_{\rm i}}{\Omega} - 1\right) \ .
    \end{aligned}
\label{eq:rad_drift2}
\end{equation}
If the angular speed of the infalling material is greater than the angular speed of the disc material then we would expect outward movement of the disc material, where the reverse is true if the angular speed of the infalling material is less than the angular speed of the disc.

The mass flow rate or accretion rate, $\dot{M}_a $, within the disc has the form:
\begin{equation}
      \begin{aligned}
    \dot{M}_a &= 2 \pi r \Sigma v_{\rm r} = \frac{2 \pi }{\frac{\partial (r^2 \Omega)}{\partial r}}\dot{\Sigma}_{\rm i} r^3(\Omega_{\rm i} - \Omega) \\
    &= 4 \pi r^2 \dot{\Sigma}_{\rm i} \left(\frac{\Omega_{\rm i}}{\Omega} - 1\right) \\
    &\approx 3.4\times10^{-10} \text{M}_\odot/\text{yr}\left(\frac{r}{1 \text{au}} \right)^2 \left(\frac{\dot{\Sigma}_{\rm i}}{10^{-7}\Sigma/\text{yr}} \right)\left(\frac{\Omega_{\rm i}}{\Omega} - 1\right) \ .
    \end{aligned}
    \label{eq:Mass_flow_rate}
\end{equation}
The characteristic flow rate of mass onto the protostar is small relative to what is observed. However, the mass flow rate is directly proportional to surface density rate of infalling material: $\dot{\Sigma}_{\rm i} $. If the infalling material is distributed inhomogeneously across the disc and $\dot{\Sigma}_{\rm i} $ has a local value that is relatively high, then $ \dot{M}_a$ is large and a local disc gap may form. Conversely, if $\dot{\Sigma}_{\rm i} $ has a local value that is small, then $ \dot{M}_a$ is relatively small and a disc ring may arise at that radial distance in the disc.

As the infalling material impacts the disc, it loses kinetic energy, which can be dissipated as radiation at the disc surface. So the total rate of energy dissipated per unit area, $D_{\rm i}$ by both sides of the disc, will be:
\begin{equation}
      \begin{aligned}
    D_{\rm i} &\approx \frac{1}{2} \dot{\Sigma}_{\rm i}\left( u_{\phi {\rm i}} - u_\phi \right)^2 
    +\frac{1}{2} \dot{\Sigma}_{\rm i} u_{\rm ri}^2  \\
    &= \frac{1}{2} \dot{\Sigma}_{\rm i} r^2 \Omega^2 \left(\frac{\Omega_{\rm i}}{\Omega} - 1\right)^2 +\frac{1}{2} \dot{\Sigma}_{\rm i} u_{\rm ri}^2 \ ,
    \end{aligned}
\end{equation}
with $u_{\phi {\rm i}}=r\Omega_{\rm i} $,  $u_\phi =r\Omega$, and $u_{\rm ri}(r)$ is the radial speed of the infalling material as it impacts the disc at a distance $r$ from the protostar. A lower bound on the energy dissipated can be obtained if we set $u_{\rm ri} \approx 0$ and assume Keplerian disc rotation,
\begin{equation}
      \begin{aligned}
    D_{\rm i} &\approx \frac{1}{2} \dot{\Sigma}_{\rm i} r^2 \Omega^2 \left(\frac{\Omega_{\rm i}}{\Omega} - 1\right)^2 
    = \frac{{\rm G}M_*\dot{M}_a }{8\pi r^3}\left(\frac{\Omega_{\rm i}}{\Omega} - 1\right)\\
    &\approx 0.01 \ \text{J}\text{m}^{-2} \text{s}^{-1} 
    \left(\frac{(M_* / \text{M}_\odot)(\dot{M}_a/10^{-10} \ \text{M}_\odot/{\rm yr})}{(r/\text{au})^3} \right) \\
    & \times \left(\frac{\Omega_{\rm i}}{\Omega} - 1\right),
    \end{aligned}
    \label{eq:Di_2}
\end{equation}
The energy dissipation given in equation (\ref{eq:Di_2}) is approximately one third of the energy dissipation for a standard viscous accretion disc \citep{2002apa..book.....F}. The energy dissipated by one side of the disc will be half the above number.

Finally, the time evolution of the disc surface density is obtained by combining equations (\ref{eq:mass_con}) and (\ref{eq:rad_drift2}):
\begin{equation}
      \begin{aligned}
      \frac{\partial \Sigma}{\partial t} &= -2 r\frac{\partial \dot{\Sigma}_{\rm i}}{\partial r}\left(\frac{\Omega_{\rm i}}{\Omega} - 1\right)- 4 \dot{\Sigma}_{\rm i} \left(\frac{\Omega_{\rm i}}{\Omega} - 1\right) \\
      & - 2 r \dot{\Sigma}_{\rm i} \left(\frac{1}{\Omega}\frac{\partial \Omega_{\rm i}}{\partial r} - \frac{\Omega_{\rm i}}{\Omega^2}\frac{\partial \Omega}{\partial r}\right) + \dot{\Sigma}_{\rm i} \ .
      \end{aligned}
      \label{eq:dSigmadt}
\end{equation}

In this paper, we set the angular speed of the disc gas to be approximately Keplerian:
\begin{equation}
    \Omega(r) \approx \sqrt{\frac{{\rm G}M_*}{r^3}} \ .
        \label{eq:Omega_K}
\end{equation}
As derived in Appendix C2 of \citet{2020MNRAS.493.4022L}, the angular speed of the infalling material that has been initially ejected from the inner regions, assuming Keplerian motion, of a protostellar disc is:
\begin{equation}
    \Omega_{\rm i}(r) \approx \frac{\ell}{r^2} \ ,
    \label{eq:Omega_{ i}}
\end{equation}
where $ \ell$ is the specific angular momentum of the particle. For the case where a particle is launched from the distance, $ r_{\rm i}$, then
\begin{equation}
    \ell \approx \sqrt{{\rm G}M_*r_{\rm i}} \ .
\end{equation}
If the particles are launched from a jet flow, then $r_{\rm i}$ is assumed to be located in the innermost regions of the disc.  \citet{2023ApJ...945L...7K} have determined that 0.3 to 0.7 au is a  realistic value for $r_{\rm i}$ - at least for the formation and initial ejection distance for micron-sized forsterite grains from the protostar Ex Lup. However, in this paper, we decided to adopt $r_{\rm i}$ = 0.05 au, to act as an extreme endpoint.

It is also possible that the angular speed of the particle may be super-Keplerian due to the "Propeller" effect of the stellar magnetosphere extending further than the co-rotation radius. In this case, the angular speed of the magnetosphere is greater than the angular speed of the disc gas. Such a situation may produce an outflow where the gas flow has super-Keplerian angular speed \citep{1997ASPC..121..241L}. 

If the infalling material arises from outside the protostellar system (i.e., the protostar and protostellar disc), then $\Omega_{\rm i}$ will be set by additional factors external to the protostellar system.

\section{Disc Ring and Gap Formation}
\label{sec:Ring/Gaps}

Gaps and rings in protostellar discs are quite common. To quote, \citet{2019ApJ...872..112V}:  ``ringlike structures are found across the full ranges of spectral type, luminosity, and age, ranging from 0.4 Myr up to 10 Myr''. 

There have been many interesting proposals for how such gaps and rings could form. The formation of planets is the most common and reasonable explanation for these phenomena (ibid.). However, the observation that rings and gaps appear in discs that are less than 0.5 Myr old \citep{2020Natur.586..228S} is a challenge for planet formation theories and suggest an additional process may be required to produce gaps and rings plus enhance the probability of planet formation.

In this section, we show that gaps and rings can form relatively quickly due to the infall of low angular momentum material onto the disc. Gap formation due to infall produces over-dense regions in the outer perimeters of the resulting rings. The inner perimeters of the resulting rings may also function as dust traps. The resulting over-dense regions may provide fertile conditions for planet formation. 

To illustrate ring and gap formation in a protostellar disc, we consider the case where the angular momentum of the infalling material is close to zero ($\Omega_{\rm i} \approx 0$ or $r_i \ll r $), and the rate of infall,  $\dot{\Sigma}_{\rm i}$, is approximately constant in time and space over a finite region of the disc. This situation may arise, for example, when an outflow is ejecting material which subsequently returns to discrete outer regions of the disc over a limited period of time. Alternatively, we may have the infall of low angular material from any external source where the infalling material lands on a specific section of the protostellar disc (e.g., \citet{2022ApJ...928...92K}).

As such, equation (\ref{eq:rad_drift2}) becomes
\begin{equation}
    v_{\rm r} = \frac{dr}{dt} = -2r\frac{\dot{\Sigma}_{\rm i}}{\Sigma} \ ,
    \label{eq:rad_drift4}
\end{equation}
while equation (\ref{eq:dSigmadt}) has the form
\begin{equation}
      \frac{\partial \Sigma}{\partial t} = 5 \dot{\Sigma}_{\rm i} \ .
      \label{eq:dSigmadt3}
\end{equation}
These equations were first solved, semi-analytically, in \citet{2017A&A...602A..52W}, who analysed the behaviour of accretion discs undergoing time-variable, face-on, infall accretion.

As discussed in \ref{sec:analytic solution}, the analytic solutions to these equations, assuming a constant rate of infalling (zero angular momentum) material onto a section of the disc, are:
\begin{equation}
    r(t) = \frac{r(t_0)}{\left( 1 + (t-t_0)/\tau_{\rm g} \right)^{2/5}} \ ,
    \label{eq:r(t)}
\end{equation}
and
\begin{equation}
    \Sigma(r(t),t) = \Sigma(r(t_0), t_0)\left( 1 + (t-t_0)/\tau_{\rm g} \right) \ ,
    \label{eq:Sigma(t)}
\end{equation}
where $t_0$ is the initial time and $\tau_{\rm g}$ is the timescale for radial movement in the disc gap due to infalling material:
\begin{equation}
    \begin{aligned}
    \tau_{\rm g} &= \frac{\Sigma(r(t_0), t_0)}{5\dot{\Sigma}_{\rm i}} \\
    &\approx 2\times10^5  {\rm yr} \ \frac{\left( \frac{\Sigma(r(t_0), t_0) } {10 \ {\rm kg \ m^{-2}}} \right)}{\left( \frac{\dot{\Sigma}_{\rm i}}{10^{-5} \ {\rm kg \ m^{-2} yr^{-1}}} \right)} \ .
    \end{aligned}
    \label{eq:tau_d}
\end{equation} 
The scaling in equation (\ref{eq:tau_d}) for $\Sigma(r(t_0), t_0)$ is guided by Hayashi's disc surface density \citep{1981PThPS..70...35H} for a minimum mass solar protoplanetary disc at a distance of 100 au from the protostar. The value for $\dot{\Sigma}_{\rm i}$ assumes a disc mass accretion rate onto the protostar of $5\times 10^{-7} \ {\rm M}_\odot {\rm yr^{-1}} $ of which 10 \% is ejected by the outflow and 1 \% of the outflowing material (i.e.,  $5\times 10^{-10} \ {\rm M}_\odot {\rm yr^{-1}} $ ) is returned to the disc between a distance of 100 au to 110 au from the protostar. It would appear that gap formation timescales in the range of $10^5$ to $10^6$ years are plausible.

As infall progresses, the radial distance of a particular region of the disc decreases and its mass surface density increases with time. In particular, when $(t-t_0) = \tau_{\rm g} $ then $r(\tau_{\rm g}) = r(t_0)/2^{2/5} \approx 0.758 r(t_0) $ and $ \Sigma(\tau_{\rm g}) = 2 \Sigma(r(t_0), t_0)$. So the radial distance shrinks by about a quarter and the surface mass density doubles.

\begin{figure}
\centering
\includegraphics[width=\textwidth]{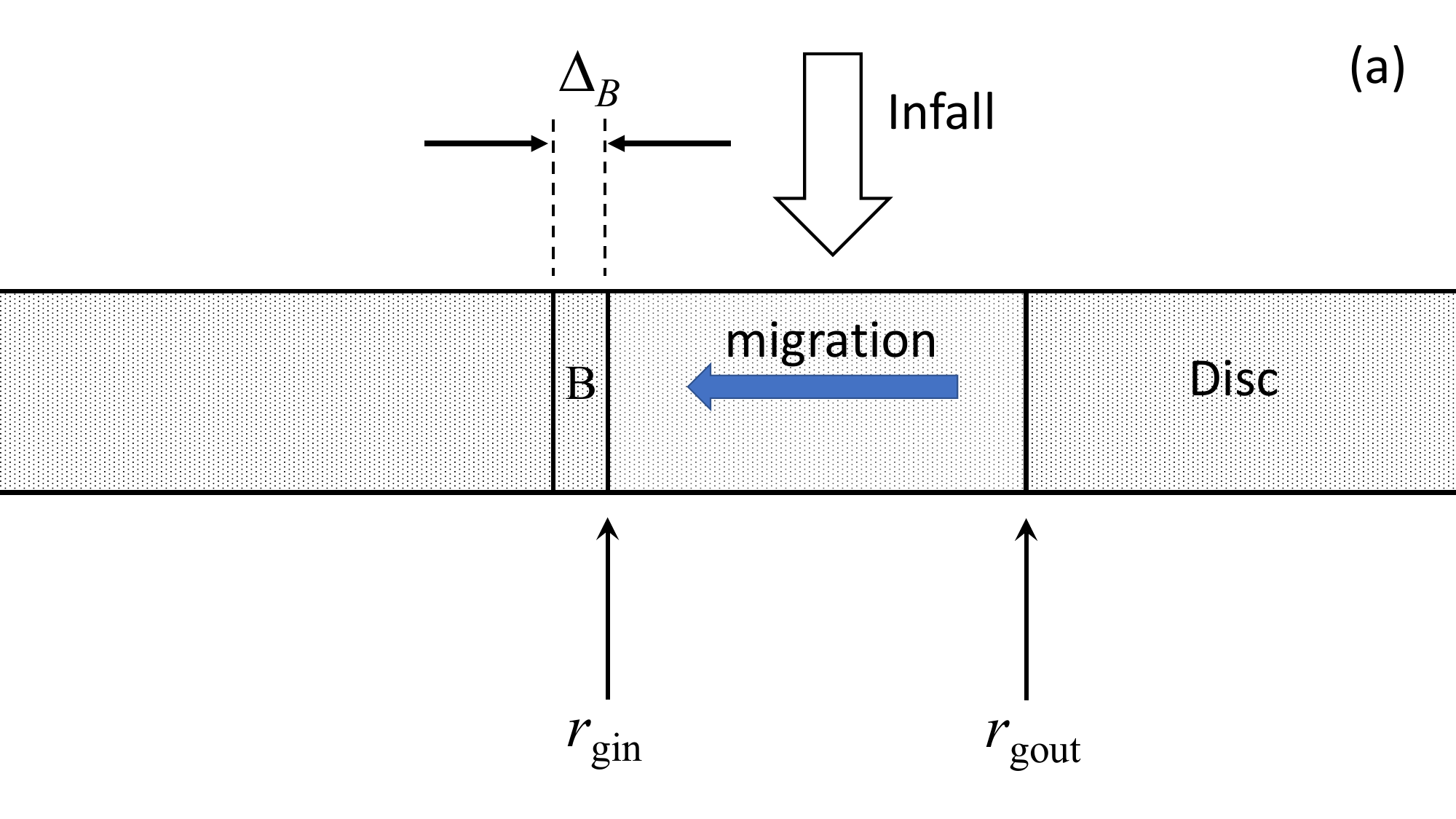}
\includegraphics[width=\textwidth]{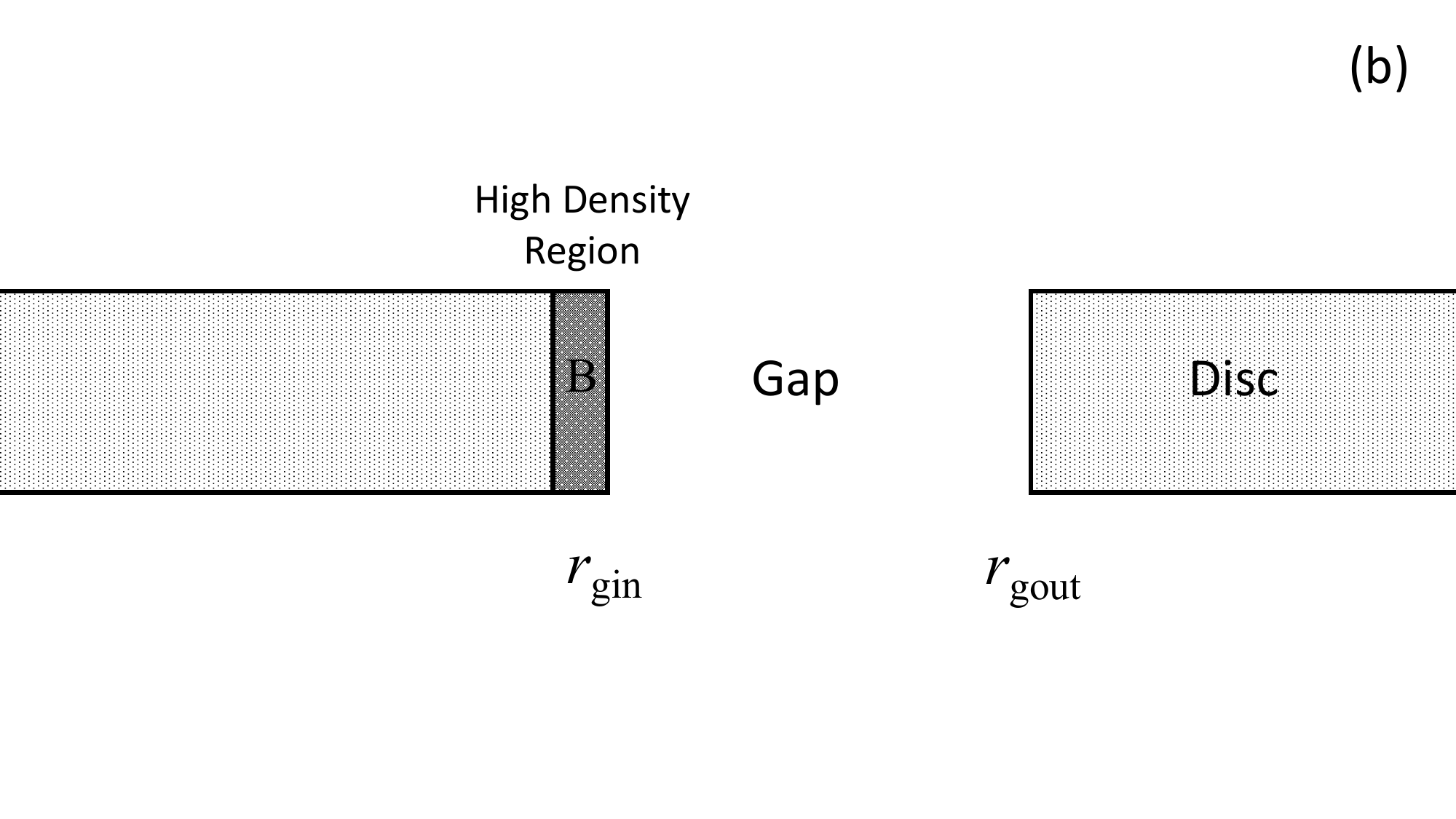}
\caption{(a) Infalling material rains down on a section of a disc, thereby inducing radial migration of material into a boundary layer B of thickness $
\Delta_{\rm B}$.(b) If the infall proceeds for a sufficiently long period of time, a gap, or a low density region may form with a high density region or pressure maximum region B at the inner edge of the gap.}
\label{fig:Grenzabschnitt}
\end{figure}

Suppose we have material raining down on a disc section, between $r_{\rm gin}$ and $r_{\rm gout}$, where the adjacent disc regions suffer relatively little or no infalling material. This affected disc region will tend to radially migrate to an inner boundary section of the disc, $B$, with thickness, $\Delta_{\rm B}$ and thereby produce a gap in the disc. Over time, this disc boundary section may develop a significantly enhanced density (Figure \ref{fig:Grenzabschnitt}).

\begin{figure}
\centering
\includegraphics[width=\textwidth]{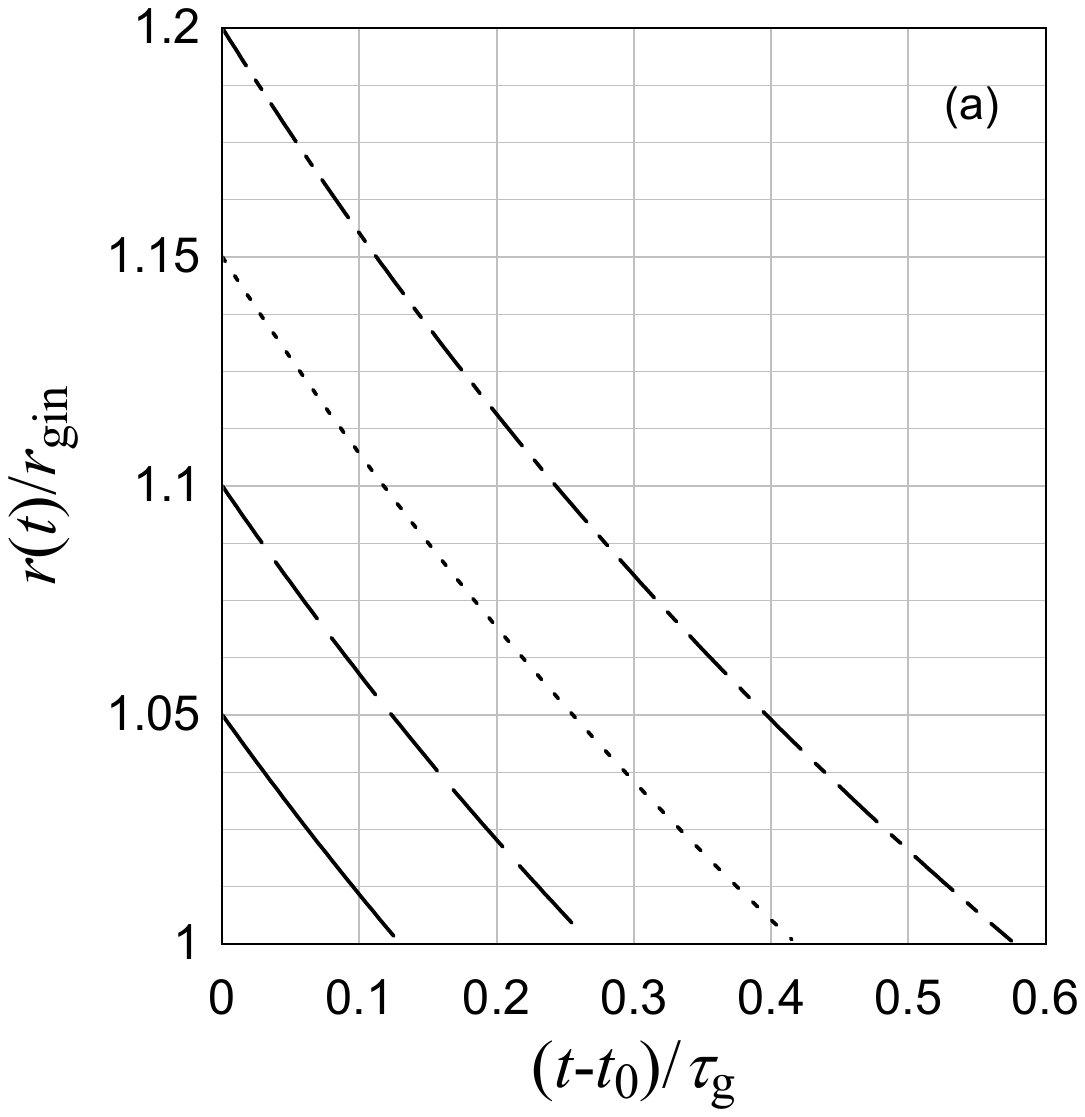}
\includegraphics[width=\textwidth]{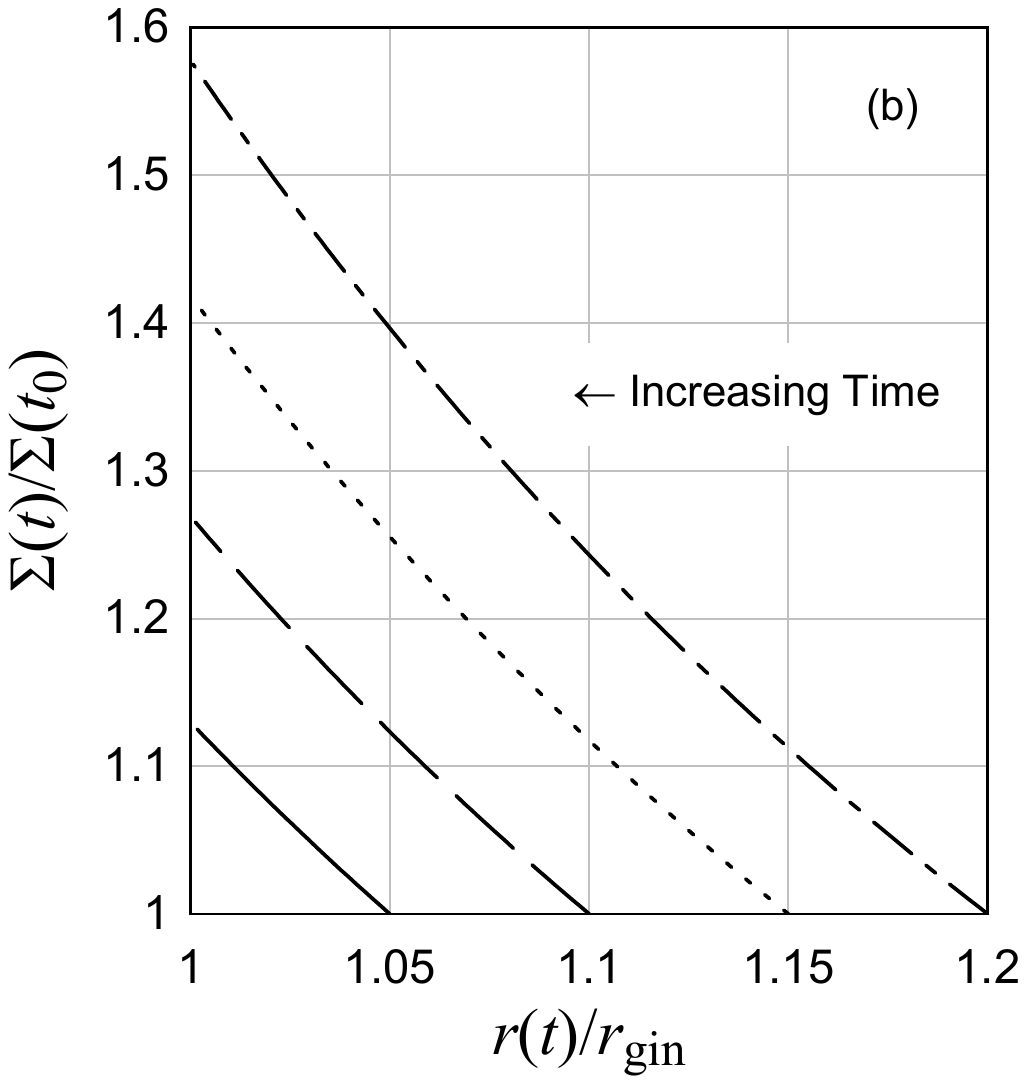}
\caption{(a) Disc radius as a function of time. As an example, the top line starts at $1.2 r_{\rm gin}$ and after a time of around 0.58$\tau_{\rm g}$ this section of the contracting disc reaches the inner gap radius ($r_{\rm gin}$).(b) The change in surface density as the gap forms. For example, the top line starts at a distance of $1.2 r_{\rm gin}$. As the gap forms, the disc radius contracts towards the inner gap radius and the disc surface density increases to near 1.6 times the original disc surface density.}
\label{fig:gap_movement}
\end{figure}

The behaviour of equations (\ref{eq:r(t)}) and (\ref{eq:Sigma(t)}) in producing the disc gap is displayed in Figure \ref{fig:gap_movement}. In Figure \ref{fig:gap_movement}(a), the lines follow the position in the disc that is undergoing infall of material. This part of the disc moves towards the inner radius of the disc gap over time. For example, the dashed line starts at 1.1 times the inner edge of the gap (1.1 $r_{\rm gin}$) and after approximately a time of 0.3 $\tau_{\rm g}$, this portion of the disc reaches the inner edge of the gap. For Figure \ref{fig:gap_movement}(b), increasing time is now right to left. We can again start at a distance at 1.1 times the inner edge of the gap. At the end of the disc migration, the surface density of the disc is approaching 1.3 times the original disc surface density.

From equation (\ref{eq:Mass_flow_rate}) the mass in the boundary section, $M_{\rm B}$, is
\begin{equation}
    \begin{aligned}
    M_{\rm B}(t) & \approx -4 \pi r_{\rm gin}^2 \dot{\Sigma}_{\rm i} \left(\frac{\Omega_{\rm i}}{\Omega} - 1\right)(t-t_0) + M_{\rm B}(t_0) \\
    & \approx 4 \pi r_{\rm gin}^2 \dot{\Sigma}_{\rm i} (t-t_0) + M_{\rm B}(t_0) 
    \end{aligned}
    \label{eq:MB(t)}
\end{equation}
 
\begin{figure}
\centering
\includegraphics[width=\textwidth]{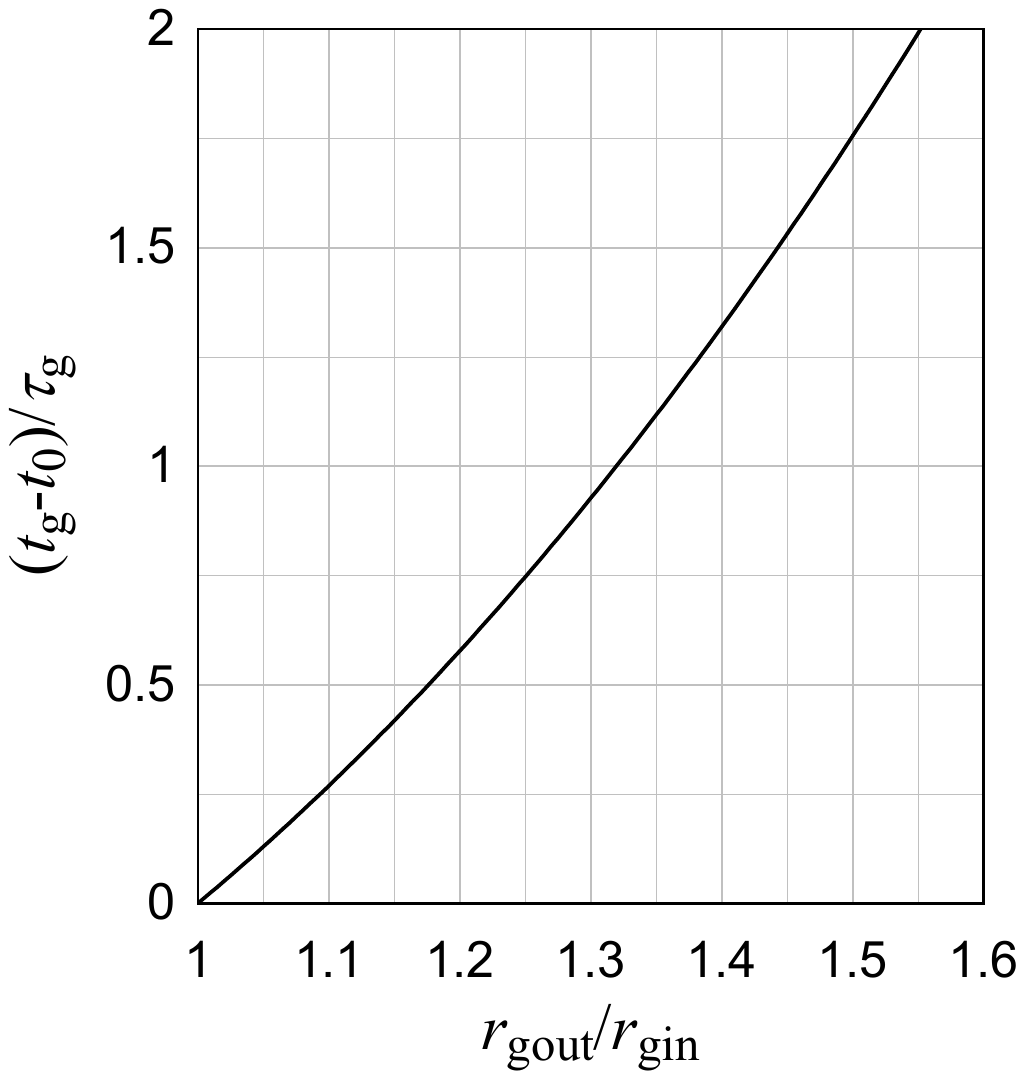}
\caption{Gap formation time as a function of gap width, where  $t_{\rm g}$ is the total time for the gap to appear, $\tau_{\rm g}$ is the timescale for disc gap formation, while $r_{\rm gout}$ and $r_{\rm gin}$ are, respectively, the outer and inner radii of the gap }
\label{fig:gap_formation_time}
\end{figure}

 The time it takes for a particular point in the nascent disc gap to move from $r(t_0)$ to $r(t)$ can be deduced from equation (\ref{eq:r(t)}). 
 
 \begin{equation}
    t-t_0 = \tau_{\rm g}\left(  \left(\frac{r(t_0)}{r(t)}\right)^{5/2} - 1 \right) \ .
    \label{eq:r(t) time}
\end{equation}
 Assuming an approximately initially similar disc surface density $\Sigma(t_0)$ in the gap region, the total time, $t_{\rm g}$, for the gap to appear is:
\begin{equation}
    \begin{aligned}
    t_{\rm g}-t_0 &=  \tau_{\rm g}\left(  \left(\frac{r_{\rm gout}}{r_{\rm gin}}\right)^{5/2} - 1 \right) \\
    &\approx \frac{\Sigma(r_{\rm gout}, t_0)}{5\dot{\Sigma_{\rm i}}}\left(  \left(\frac{r_{\rm gout}}{r_{\rm gin}}\right)^{5/2} - 1 \right) \ .
    \end{aligned}
    \label{eq:rgaptime}
\end{equation}
A plot of equation (\ref{eq:rgaptime}) is given in Figure \ref{fig:gap_formation_time}. As an example, to understand this graph, if the inner and outer gap radii are 100 au and  130 au, respectively then the ratio of the two radii is 1.3.  Consequently, it will approximately take a time $ \tau_{\rm g}$ for such a gap to form.

 Substituting equation (\ref{eq:rgaptime}) into equation (\ref{eq:MB(t)}) gives the total amount of mass accreted into the boundary region, $B$:
\begin{equation}
    M_{\rm B}(t_{\rm g}) \approx \frac{4}{5} \pi r_{\rm gin}^2 \Sigma(r_{\rm gout}, t_0)\left(  \left(\frac{r_{\rm gout}}{r_{\rm gin}}\right)^{5/2} - 1 \right) + M_{\rm B}(t_0) 
    \label{eq:MB(t_e)}
\end{equation}

Assuming $\Delta_{\rm B} \ll r_{\rm gin}$, the area of the boundary region is approximately $2\pi r_{\rm gin} \Delta_{\rm B}$, which, combined with equation (\ref{eq:MB(t)}), gives a boundary surface density of

\begin{equation}
    \Sigma_{\rm B}(r_{\rm gin}, t)  \approx  \frac{2 r_{\rm gin} \dot{\Sigma}_{\rm i} (t-t_0)}{\Delta_{\rm B}} + \Sigma_{\rm B}(r_{\rm gin}, t_0) \ .
    \label{eq:SigmaB(t)}
\end{equation}

The gap is fully formed at $t = t_{\rm g}$ and
\begin{equation}
    \Sigma_{\rm B}(r_{\rm gin}, t_{\rm g})  \approx  \frac{2 r_{\rm gin} \Sigma(r_{\rm gout}, t_0)}{5 \Delta_{\rm B}}\left(  \left(\frac{r_{\rm gout}}{r_{\rm gin}}\right)^{5/2} - 1 \right) + \Sigma_{\rm B}(r_{\rm gin}, t_0) \ .
    \label{eq:SigmaB(te)}
\end{equation}
If $\Sigma(r_{\rm gout}, t_0) \approx \Sigma_{\rm B}(r_{\rm gin}, t_0) $ then

\begin{equation}
    \begin{aligned}
   \frac{\Sigma_{\rm B}(r_{\rm gin}, t_{\rm g})}{\Sigma_{\rm B}(r_{\rm gin}, t_0)}  &\approx  \frac{2 r_{\rm gin} }{5 \Delta_{\rm B}}\left(  \left(\frac{r_{\rm gout}}{r_{\rm gin}}\right)^{5/2} - 1 \right) + 1 \\
   &= 40 \ \frac{1}{(\Delta_{\rm B}/0.01 \ r_{\rm gin})} \left(  \left(\frac{r_{\rm gout}}{r_{\rm gin}}\right)^{5/2} - 1 \right) + 1 \ .
    \end{aligned}
    \label{eq:SigmaB(te)/SigmaB(t0)}
\end{equation}

If $\Delta_{\rm B} \ll r_{\rm gin}$ then the increase in surface density in the B (boundary) region could be quite significant. In equation (\ref{eq:SigmaB(te)/SigmaB(t0)}), we have set $\Delta_{\rm B} = 0.01 \ r_{\rm gin}$ as an example of the possible over-density of material of approximately a factor 40 in the B region of the gap. This over-density could be enhanced in the inner regions of the disc, where the ratio  $r_{\rm gout}/r_{\rm gin}$ may be relatively large. Such a density enhancement is also a local pressure maximum and would act as a dust/particle trap thereby increasing the possibility of planetesimal/planet formation. 

So, the earliest stages of planet formation may arise on the proximal (i.e., closer to the protostar) edge of a disc gap. However, the distal (i.e., the farthest) edge of the resulting gap will also act as a pressure maximum and dust trap. Either way, infall may produce a gap in the disc where the edges of the resulting gap are prime sites for planet formation.

The formation of gaps by this infall process would essentially be semi-random in terms of radial distance from the protostar. If we ignore the, possibly important, influence of gas drag above the disc, the ballistic path of a particle that is ejected from a disc wind/jet is strongly dependent on the initial vertical (i.e., perpendicular to the midplane) speed of the particle \citep{1995Icar..116..275L,2020MNRAS.493.4022L}. For a system of particles with a mean size and density, a "high" jet speed could eject the particles from a protostellar system or to the outer regions of a protostellar disc. A "low" jet speed would transport particles to the more inner regions of the disc.  The resulting radial position of gap would be dependent on random fluctuations in the mean jet/wind speed.

\subsection{Enriched Concentration of Processed Dust and Pebbles}
\label{subsec:Enrichment}

In the outflow picture of dust infall, dust and pebbles from the inner disc are ejected by an outflow and/or disc wind. This material is observed to be processed by stellar radiation and/or thermal activity in the inner disc. As the disc gap increases in size, the concentration of processed material in the contracting disc between $r_{\rm gin}$ and $r_{\rm gout}$ will increase, as will the concentration of processed material in the B region (Figure \Ref{fig:Grenzabschnitt}). 

The mass of original material (i.e., dust and gas), $M_{\rm G}$ in the gap between $r_{\rm gin}$ and $r_{\rm gout}$ is approximately
\begin{equation}
    M_{\rm G} \approx \Sigma_{\rm G}\pi \left(r_{\rm gout}^2 - r_{\rm gin}^2 \right) \ ,
\end{equation}
where $\Sigma_{\rm G}$ is the disc surface density in the gap - which, in this case, is assumed to be approximately constant between $r_{\rm gin}$ and $r_{\rm gout}$.
 After the gap is formed, the mass of the material at B is given by equation (\ref{eq:MB(t_e)}). If we assume that the boundary layer B is sufficiently small, we can ignore the initial amount of material in the B region: $ M_{\rm B}(t_0)$, so the mass of processed infall dust and pebbles at B, $M_{\rm Bi}$, is
 
\begin{equation}
    M_{\rm Bi} \approx M_{\rm B}(t_{\rm g}) - M_{\rm B}(t_0) - M_{\rm G}
    \label{eq:MBi}
\end{equation}

The ratio of processed dust and pebbles at B to the original mass of dust in the gap, $R_{\rm BiG}$. is
\begin{equation}
    \begin{aligned}
   R_{\rm BiG} & = \frac{M_{\rm Bi}}{f_{\rm dg}M_{\rm G}}  
   \approx \frac{M_{\rm B}(t_{\rm g}) - M_{\rm B}(t_0) - M_{\rm G}}{f_{\rm dg}M_{\rm G}} \\
    & = \frac{1}{f_{\rm dg}} \left( \frac{4\left(\left(\frac{r_{\rm gout}}{r_{\rm gin}}\right)^{5/2} - 1\right) }{5\left(\left(\frac{r_{\rm gout}}{r_{\rm gin}}\right)^{2} - 1\right)}- 1\right) \ ,
    \end{aligned}
    \label{eq:MBi/fdgMG}
\end{equation}
where $f_{\rm dg} \approx 0.01$ is the dust to gas mass ratio in the disc.

\begin{figure}
\centering
\includegraphics[width=\textwidth]{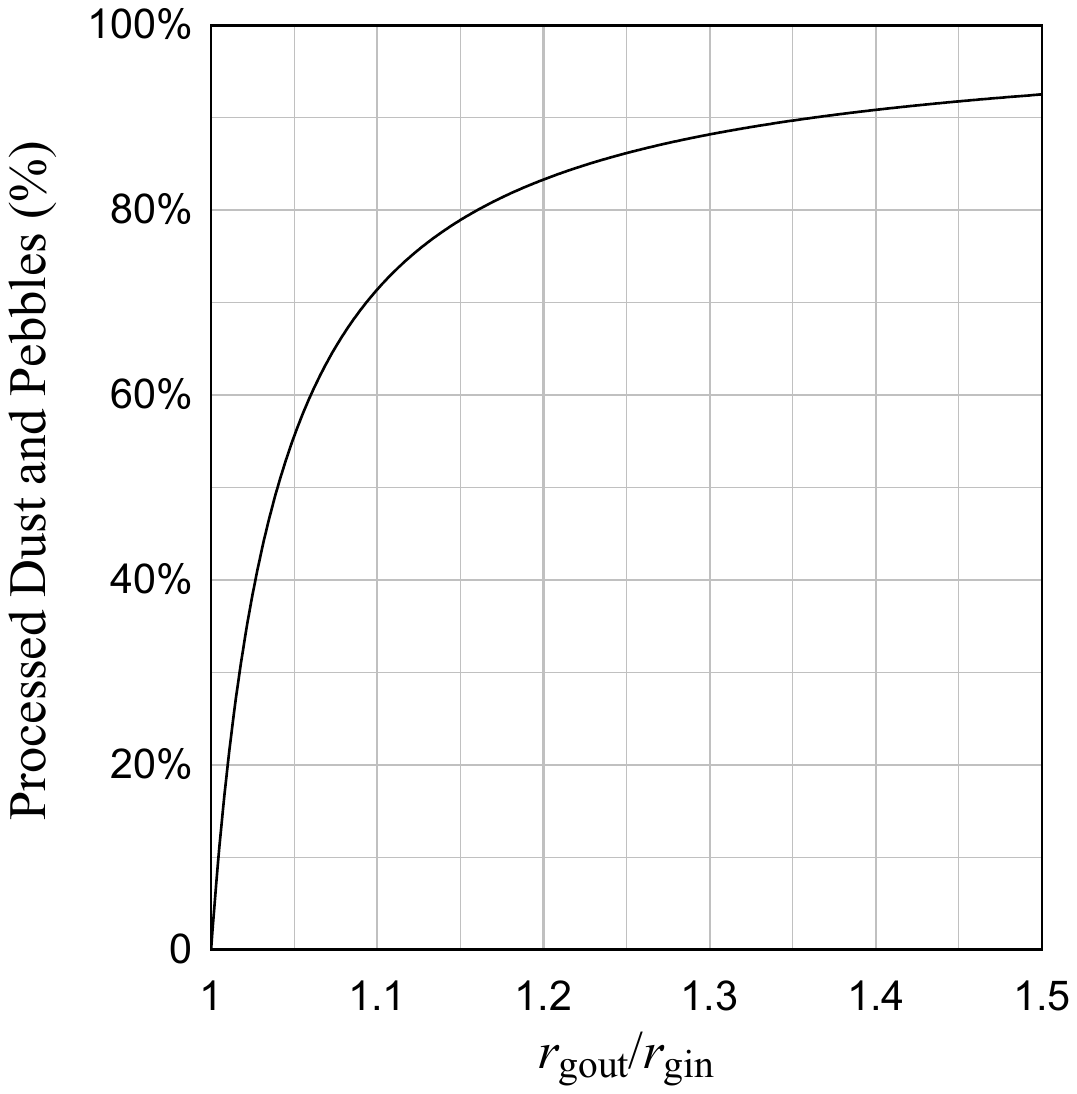}
\caption{The mass enrichment in the boundary layer B of processed infall dust and pebbles relative to the original dust in the gap, as a function of the ratio of the outer gap distance to the inner gap distance: $r_{\rm gout}/r_{\rm gin}$. If this gap ratio is, for example, 1.2 then approximately 83\% of the material in the B region is composed of processed dust and pebbles.}
\label{fig:Dust_Enrichment}
\end{figure}

The enrichment of processed dust and pebbles at B: 
\begin{equation}
    \frac{M_{\rm Bi}}{f_{\rm dg}M_{\rm G} + M_{\rm Bi}} \ .
\end{equation}
 as a function of the gap distance ratio: $r_{\rm gout}/r_{\rm gin}$ is shown in Figure (\ref{fig:Dust_Enrichment}). Here we can see that a gap distance ratio of 1.1 would imply that approximately 70\% of the material in the B region was composed of processed dust and pebbles. Such a result may have application in the early solar system: if plantesimals formed from "B" region material then most of the material we observe in primitive meteorites and comets may have been preprocessed dust and pebbles. 

\subsection{Dust Traps}
\label{subsec:Multi Gap Dust Trap}

A difficulty in the scenario that is schematically depicted in Figure \ref{fig:Grenzabschnitt} is that the concentrated dust in the boundary "B" region may simply move out of the region due to radial drif                           t on timescales that are shorter then the formation timescale of the "B" region. This would not be an opportune situation for planet formation. As noted previously, the outer edge of the gap in Figure \ref{fig:Grenzabschnitt} will naturally act as a dust drap, due to the increase in pressure gradient between the gap and the disc.

\begin{figure}
\centering
\includegraphics[width=\textwidth]{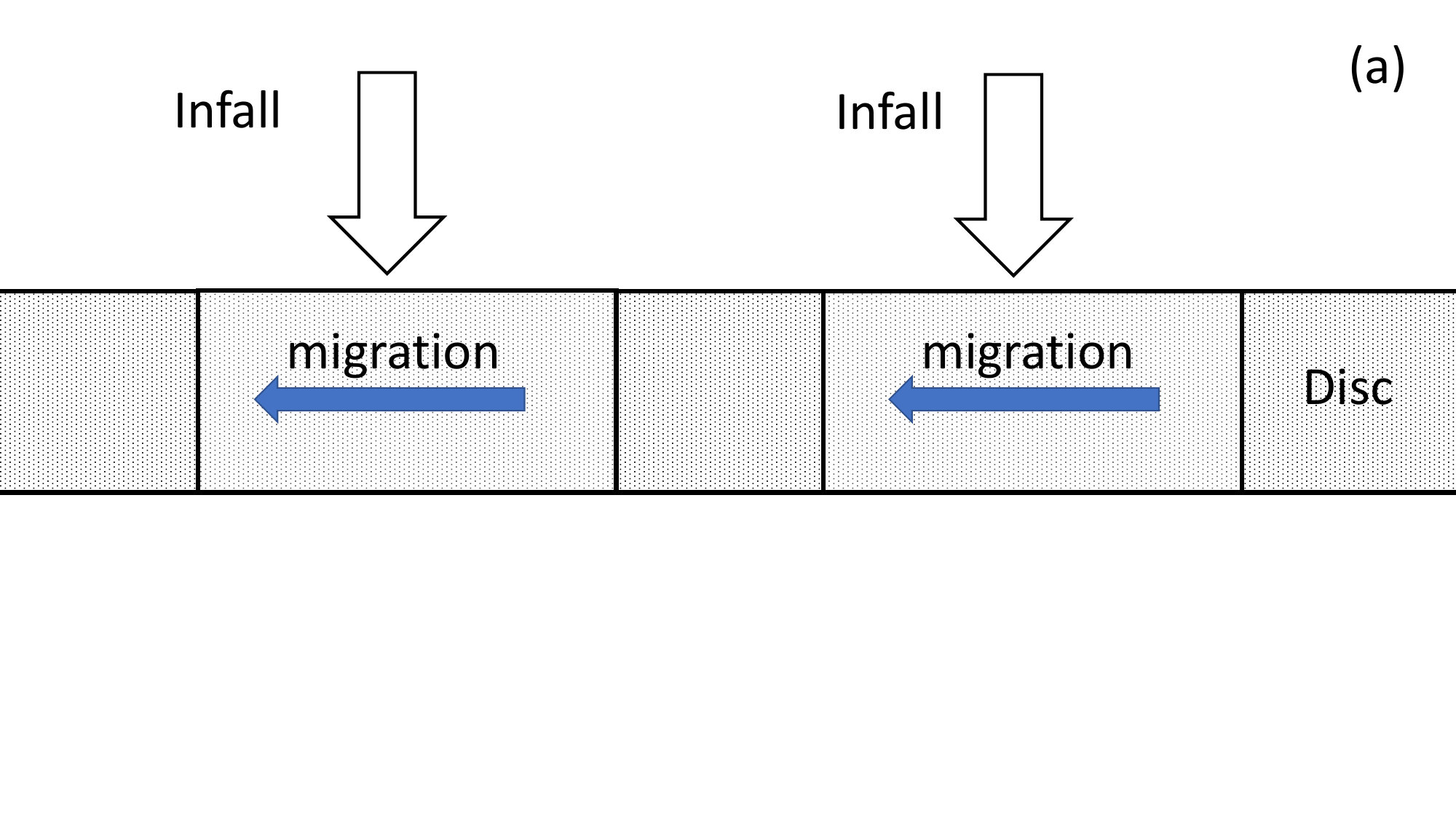}
\includegraphics[width=\textwidth]{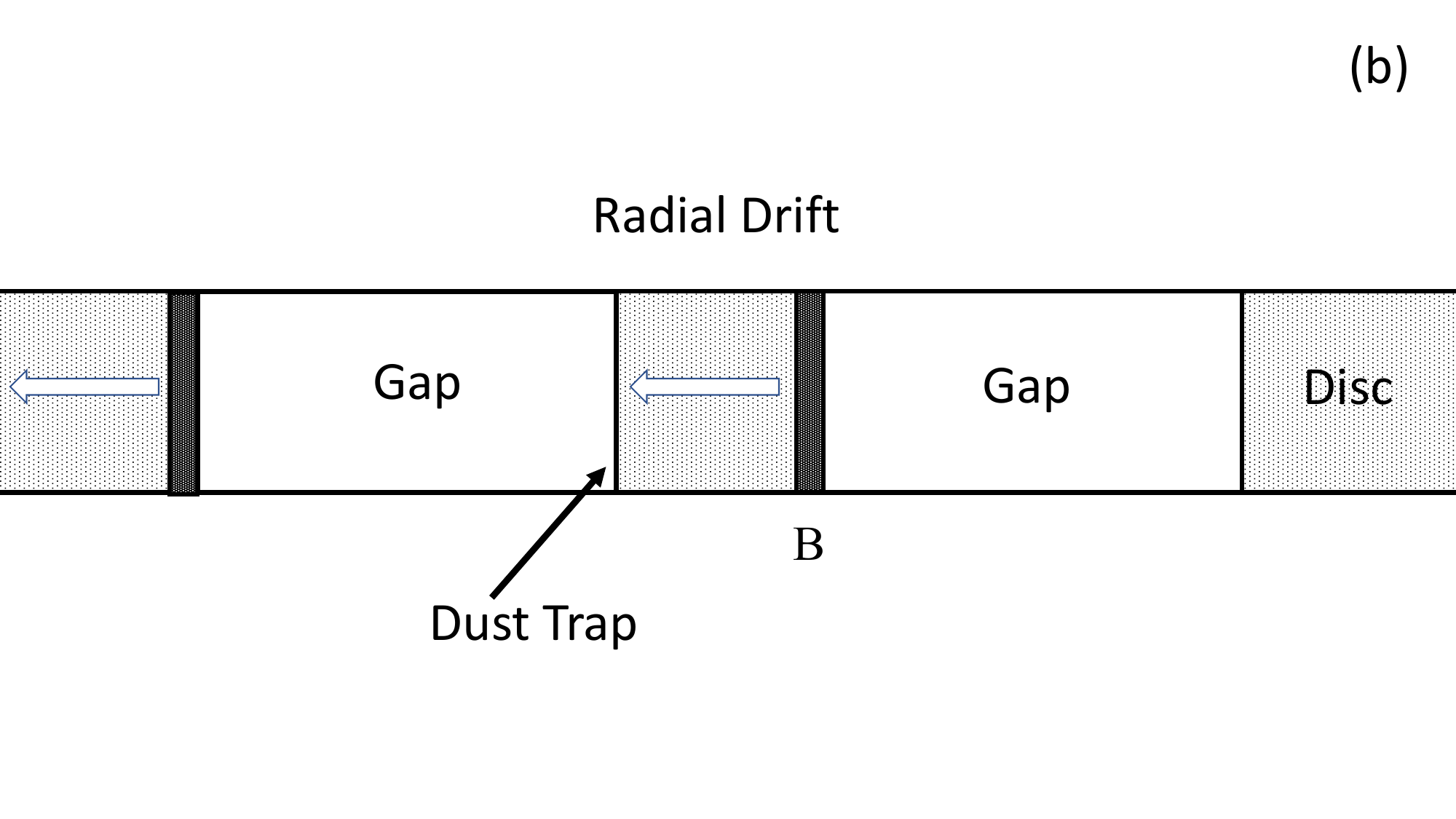}
\includegraphics[width=\textwidth]{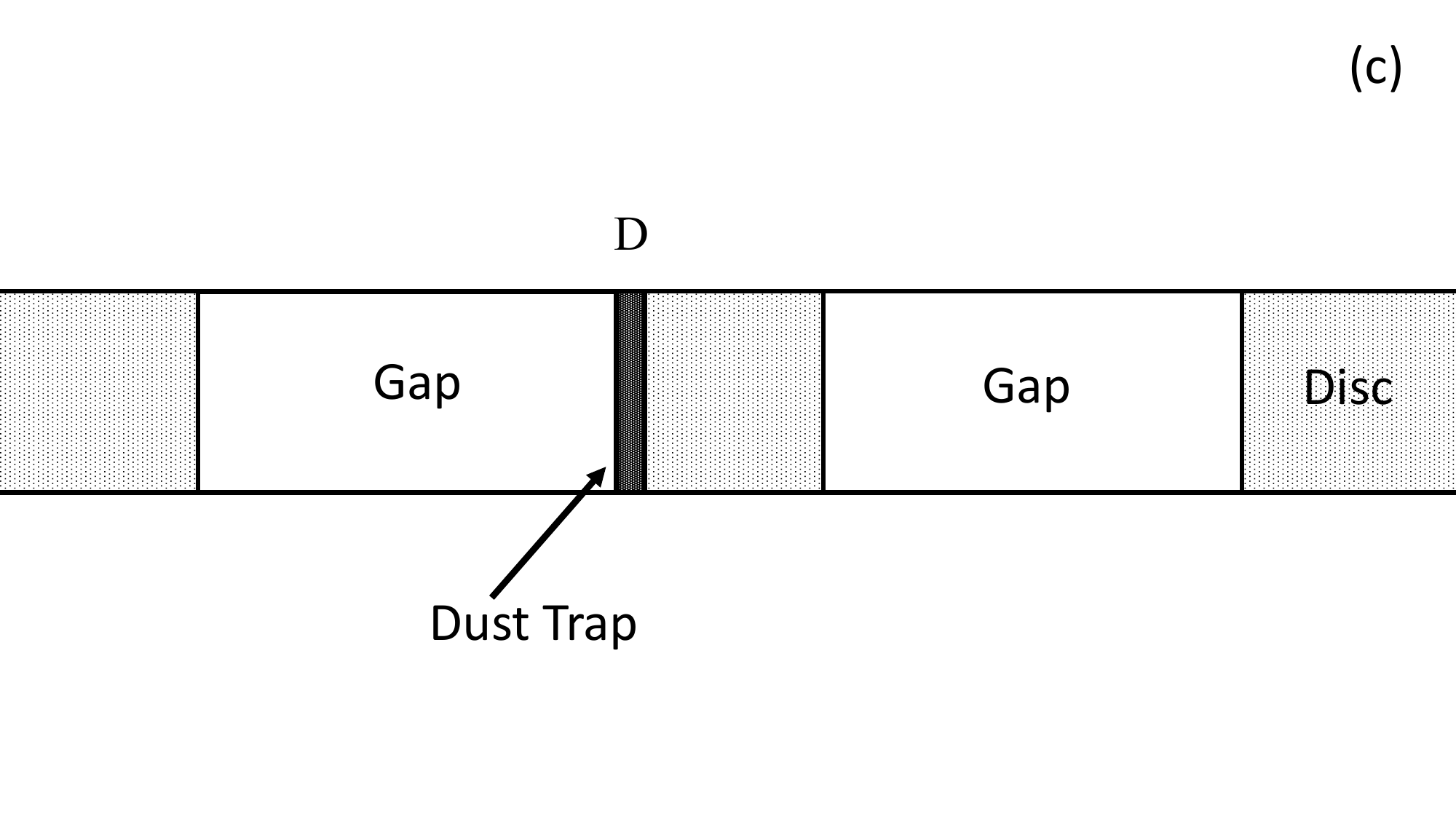}
\caption{(a) Infalling material rains down on two separate sections of a disc, thereby inducing radial migration of material into boundary layers located at the inner gap edges. (b) If the infall proceeds for a sufficiently long period of time, gaps, or low density regions may form with a high density region or pressure maximum region B at the inner edge of the gaps. During the formation f these high density regions, the particulate material may radially drift towards the protostar. (c) Particulate material is subsequently trapped in the D (dust trap) region which naturally forms at the outer edge of the inner gap. Both the B and D regions may have an over abundance of particulate material that is advantageous for planetesimal formation. }
\label{fig:GrenzabschnittII}
\end{figure}

So, another possibility is that the infall of material forms two gaps instead of one as is illustrated in Figure \ref{fig:GrenzabschnittII}. In this case, the particulate material from the outer gap can potentially be trapped by the outer edge of the inner gap. To obtain an indication of the drift time required for a particle of a certain size and density, we consider a ring of material that is created between two gaps.

From \ref{sec:drift_eqns}, the drift time for particles in a model disc is:
\begin{equation}
    t - t_0 \approx \tau_{\rm drift} \left( \left( \frac{r_0}{r}\right)^{3/4} - \left( \frac{r_0}{r_{\rm t_0}}\right)^{3/4} \right) \ ,
    \label{eq:drifttime}
\end{equation}
with $t_0$ the initial time, $r_{\rm t_0}$ the initial position of the particle, and $r_0$ = 1 au is the distance scale used in our model disc (\ref{sec:drift_eqns}).

The drift timescale $\tau_{\rm drift}$ is given by
\begin{equation}
    \tau_{\rm drift} = \frac{8 r^2_0 \mu m_{\rm H}}{ 21 \tau_{\rm s0} {\rm k}_{\rm BP} T_{\rm g0}} \approx 1.5\times10^{5}  \left( \frac{1 \ {\rm mm}}{a_{\rm p}} \right) {\rm yrs} \ ,
    \label{eq:drifttimescale}
\end{equation}
here, ${\rm k}_{\rm BP}$ the Boltzmann constant, $\mu$ the mean molecular mass ($\approx 2.3$), $m_{\rm H}$ the hydrogen atom mass, $\tau_{\rm s0}$ is the scale factor for the stopping time and $T_{\rm g0}$ is a scale factor (= 500 K) for the temperature of the gas (see \ref{sec:drift_eqns}).

Equation (\ref{eq:drifttime}) is approximately valid for a particle with a radius less than about a metre. Equation (\ref{eq:drifttimescale}) indicates that particles with radii less than about 1 mm are going to have drift times that tend to be comparable to or in excess of the gap formation time scales. This also depends, of course, on the width of the disc ring between the two gaps. Noting that caveat, particles with radii less than 0.1 mm will tend to stay in the compressed B regions, while larger particles will have a greater chance of being caught in a inter-gap dust trap D (Figure \ref{fig:GrenzabschnittII}).

\section{Discussion}

\label{sec:discussion}

There is a growing body of evidence that refractory, particulate material like Forsterite (Mg$_2$SiO$_4$) dust
is ejected from the inner disc regions of protostellar systems (possibly between 0.3 and 0.7 au for the protostellar system EX Lup) and is either completely ejected from the protostellar system (e.g., HOPS-68 \citep{2011ApJ...733L..32P}) or lands back in the disc around 3 au (for EX Lup) from the ejection point near the star \citep{2012ApJ...744..118J,2019ApJ...887..156A,2023ApJ...945L...7K}.

The physical mechanisms that eject particulate material from accretion discs have been debated over many years. These mechanisms include stellar winds \citep{1977ApJ...217..693H}, disc winds \citep{1992Icar..100..608L,1995Icar..116..275L,1996cppd.proc..285L,1996Sci...271.1545S,2010ApJ...725.1421H,2012ApJ...758..100B,2012M&PS...47.1922S,2016ApJ...821....3M,2019ApJ...882...33G}, photophoresis \citep{2005ApJ...630.1088K,2016MNRAS.458.2140C}, photoevaporation \citep{2016MNRAS.461..742H} and radiation pressure combined with disc winds \citep{2021MNRAS.500..506V}. The complexity of the physics is such that {\it all} of these processes may play a role depending on the material that is being ejected and where in the disc the material is located. Regardless of the physical mechanisms that drive this ejection process, once the particles start moving freely across the face of the accretion disc, their orbital angular momentum should become small relative to their radial position in the accretion disc. 

We have found that the infall of ejected, processed low angular momentum dust/pebbles/gas from a protostellar jet and/or disc wind onto a discrete, somewhat random region of a protostellar disc can produce a disc gap on a timescales of of $10^5$ to $10^6$ years (\S\ref{sec:Ring/Gaps}).  This may, partially, answer the question of how the observed gaps and rings can form in even the youngest protostellar systems \citep{2019ApJ...872..112V}. Infall of low angular momentum material from a nascent molecular cloud onto a protostellar system may also produce the same effect \citep{2022ApJ...928...92K}. 

The infall of low angular momentum material onto a protostellar disc can produce disc gaps, which then lead to over-dense regions on the inner or outer edges of the gaps, which encourage planet formation. Infall is the "chicken" that lays the "egg" of planet formation.

The original disc material in the gap plus the infalling material, can be compressed onto the inner edge of the gap, i.e., the "B" boundary region of Figure \ref{fig:Grenzabschnitt}, where most of the solid material in this compressed B region consists of processed dust and pebbles (Figure \ref{fig:Dust_Enrichment}). Another possibility is that mulitple gaps form, due to the non-uniform infall of material over a disc interval and that material drifts through a thin disc ring, to be trapped in the "D" dust trap region on the other side of the ring (Figure \ref{fig:GrenzabschnittII}).

In the more parochial context of the formation Solar System, this B region could contain O$^{16}$-rich material (e.g., Calcium Aluminium Inclusions (CAIs) and Amoeboid Olivine Aggregates (AOAs)) which likely formed near the protoSun/inner edge of the disc \citep{2016MNRAS.462.1137L,2020E&PSL.53516088L} plus processed, O$^{16}$-poor, forsterite dust from the inner disc regions. In this case, the CAIs/AOAs could be ejected from the interaction between the solar magnetosphere and the inner edge/region of the solar accretion disc, while the forsterite grains could be ejected via disc winds produced further from the protoSun. 

\begin{figure}
\centering
\includegraphics[width=\textwidth]{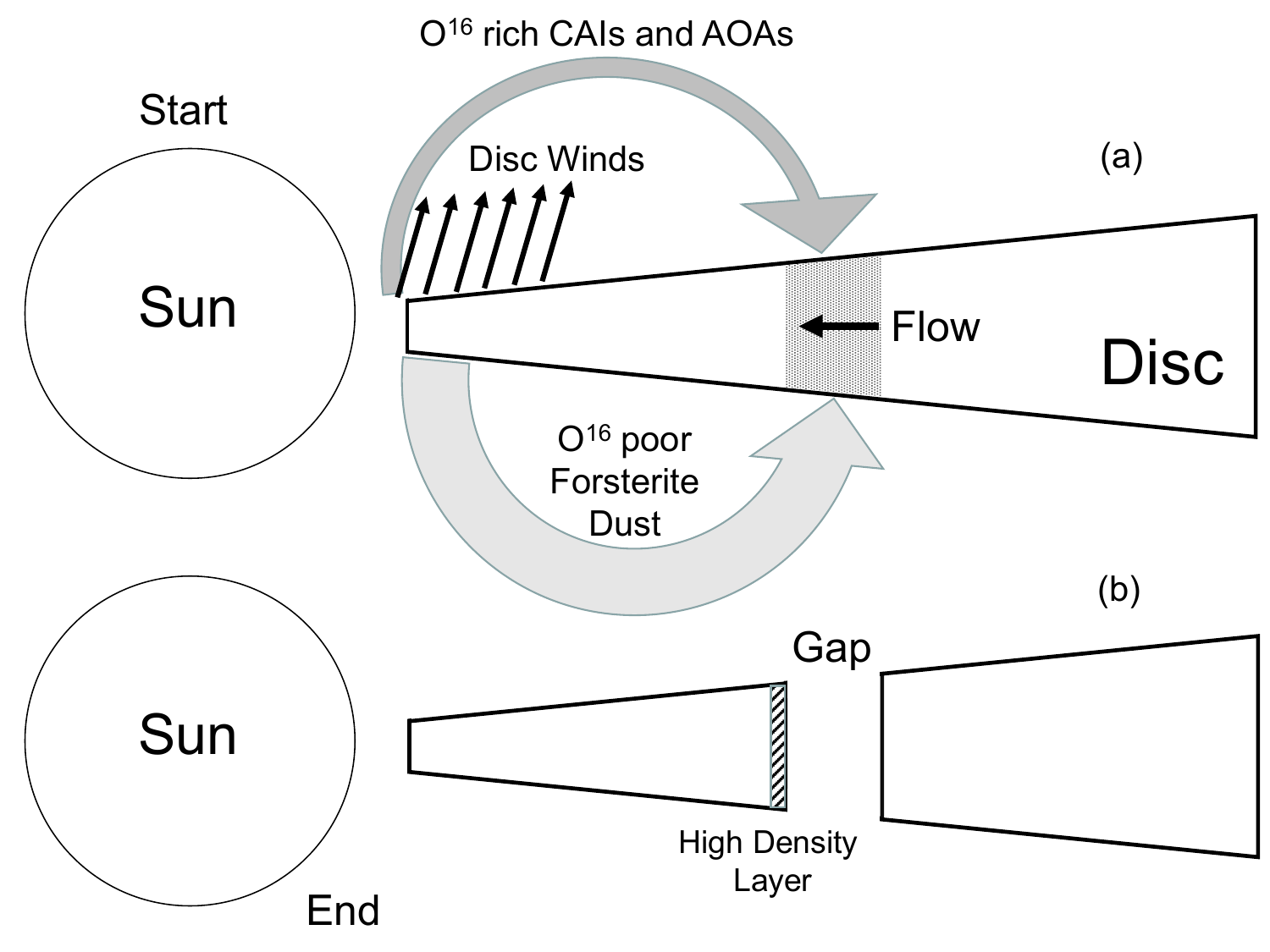}
\caption{(a) The disc gaps would be produced from processed infall material. In the context of the early Solar System, the jet flows closest to the protoSun, ejected  Calcium Aluminium Inclusions (CAIs) and Amoeboid Olivine Aggregates (AOAs) that are both enriched with O$^{16}$, where the ratio of oxygen isotopes are similar to those found in the Sun \citep{2011Sci...332.1528M}.Observations from EX Lup suggest that disc winds between 0.3 and 0.7 au from the star, eject/process amorphous silicate into crystalline forsterite that lands further away from the protoStar. In the Solar System, such forsterite would very probably be O$^{16}$ poor. (b) The subsequent gap formation would concentrate these materials to, potentially, form the base material for pre-chondritic planetesimals}
\label{fig:oxygen_mixture}
\end{figure}

The resulting gap formation and concentrating of infalling, 
processed materials would then provide the foundation for prechondritic planetesimals. Chondrules and the base materials for chondritic planetesimals might then be produced by collisions between the pre-chondritic planetesimals (Figure \ref{fig:oxygen_mixture}) . This general scenario is qualitatively consistent with the hypothetical formation process required to explain the isotopic diversity observed in carbonaceous chondrites \citep{2023ApJ...946L..34H}.

\section{Conclusions}

\label{sec:conclusions}

In this study, we have examined the general case of what happens when low angular momentum material falls onto a protostellar disc. We have used the resulting mass and angular momentum equations (equations (\ref{eq:mass_con}) and (\ref{eq:ang_con2})) to derive, for an inviscid disc, a number of quantities including the radial speed of disc material (equation (\ref{eq:rad_drift2})), the rate of mass flow in the disc (equation (\ref{eq:Mass_flow_rate})) and the time evolution of the disc surface density (equation (\ref{eq:dSigmadt})). 

We then assume that the source of this infalling low angular momentum material onto the protoplanetary disc is the ejection of, mainly, particulate material from the inner disc regions, where this ejected, relatively low angular momentum, material subsequently falls onto the outer disc regions. Such particle ejection and transport is consistent with observations and long standing theoretical predictions. Equations (\ref{eq:rad_drift2}) and (\ref{eq:dSigmadt}) are then applied to the formation of disc rings and gaps (\S \ref{sec:Ring/Gaps}). Using the derived equations, we show that the recycling of infalling material is probably not large enough to drive the observed accretion flow from a protostellar disc onto its protostar. However, infall can produce gaps/rings in discs on timescales of $10^5$ to $10^6$ years.

The formation of gaps and rings occurs when infalling, low angular momentum material lands on a discrete section of the disc instead of the entire disc surface. Such a phenomenon is consistent with observations. The affected section of the disc then moves towards the protostar and the disc material will slowly ram into the adjacent, interior disc, which is not suffering the influence of infalling material. This process may produce an over abundance of material in a compressed layer on the inner edge of the gap (\S \ref{subsec:Enrichment}). The amount of enrichment can be significant. For example, if the ratio of the radius of the outer edge to the inner edge of the gap is 1.1 then about 70\% of the solid material in the compressed layer is composed of processed dust and pebbles from the inner regions of the disc (Figure \ref{fig:Dust_Enrichment}).

Alternatively, infall may be somewhat random in space and time thereby producing multiple gaps. In such a circumstance, particulate material may drift from the inner edge of an outer gap, to the outer edge of an adjacent, inner gap (Figure \ref{fig:GrenzabschnittII}). The outer edge of a gap is a natural dust trap. Reasonable drift timescales tend to be applicable for particles greater than 0.1 mm in radius, where our analysis has only considered particles less than a metre in radius (\S \ref{subsec:Multi Gap Dust Trap}).

We suggest that gap formation due to infall lays the foundation for planetesimal formation by compressing the material in the gap to the inner edge of the gap or, via radial drift, to the adjacent outer edge of another gap. Gaps can be produced on relatively short timescales from the inception of the protostellar system. 

In our protosolar system, such a process may have collected O$^{16}$-poor forsterite dust from the inner regions of the protosolar disc and O$^{16}$-rich CAIs and AOAs from the inner edge regions of the protosolar disc, thereby constructing a region favourable to the formation of prechondritic planetesimals.

\begin{acknowledgement}
I thank Professor Sarah Maddison, Dr Geoffrey Bryan and our talented colleagues at Swinburne University's Centre for Astronomy and Supercomputing for their friendship, collaboration, and support. I also thank the anonymous reviewer for his/her constructive comments that greatly improved the manuscript.
\end{acknowledgement}

\printendnotes

\printbibliography

@ARTICLE{2011ARA&A..49...67W,
       author = {{Williams}, Jonathan P. and {Cieza}, Lucas A.},
        title = "{Protoplanetary Disks and Their Evolution}",
      journal = {\araa},
         year = 2011,
        month = sep,
       volume = {49},
       number = {1},
        pages = {67-117},
          doi = {10.1146/annurev-astro-081710-102548},
archivePrefix = {arXiv},
       eprint = {1103.0556},
 primaryClass = {astro-ph.GA},
       adsurl = {https://ui.adsabs.harvard.edu/abs/2011ARA&A..49...67W}
}

@ARTICLE{2019NewA...70....7M,
       author = {{Martin}, R.~G. and {Nixon}, C.~J. and {Pringle}, J.~E. and {Livio}, M.},
        title = "{On the physical nature of accretion disc viscosity}",
      journal = {\na},
         year = 2019,
        month = jul,
       volume = {70},
        pages = {7-11},
          doi = {10.1016/j.newast.2019.01.001},
archivePrefix = {arXiv},
       eprint = {1901.01580},
 primaryClass = {astro-ph.HE},
       adsurl = {https://ui.adsabs.harvard.edu/abs/2019NewA...70....7M}
}

@ARTICLE{1991ApJ...376..214B,
       author = {{Balbus}, Steven A. and {Hawley}, John F.},
        title = "{A Powerful Local Shear Instability in Weakly Magnetized Disks. I. Linear Analysis}",
      journal = {\apj},
         year = 1991,
        month = jul,
       volume = {376},
        pages = {214},
          doi = {10.1086/170270},
       adsurl = {https://ui.adsabs.harvard.edu/abs/1991ApJ...376..214B}
}

@INPROCEEDINGS{2019EAS....82..391F,
       author = {{Fromang}, S. and {Lesur}, G.},
        title = "{Angular momentum transport in accretion disks: a hydrodynamical perspective}",
    booktitle = {EAS Publications Series},
         year = 2019,
       series = {EAS Publications Series},
       volume = {82},
        month = jun,
        pages = {391-413},
          doi = {10.1051/eas/1982035},
       adsurl = {https://ui.adsabs.harvard.edu/abs/2019EAS....82..391F}
}

@BOOK{2002apa..book.....F,
       author = {{Frank}, Juhan and {King}, Andrew and {Raine}, Derek J.},
        title = "{Accretion Power in Astrophysics: Third Edition}",
         year = 2002,
        publisher = {Cambridge University Press},
       adsurl = {https://ui.adsabs.harvard.edu/abs/2002apa..book.....F}
}

@ARTICLE{1973A&A....24..337S,
       author = {{Shakura}, N.~I. and {Sunyaev}, R.~A.},
        title = "{Black holes in binary systems. Observational appearance.}",
      journal = {\aap},
         year = 1973,
        month = jan,
       volume = {24},
        pages = {337-355},
       adsurl = {https://ui.adsabs.harvard.edu/abs/1973A&A....24..337S}
}

@ARTICLE{2017ApJ...843..150F,
       author = {{Flaherty}, Kevin M. and {Hughes}, A. Meredith and {Rose}, Sanaea C. and {Simon}, Jacob B. and {Qi}, Chunhua and {Andrews}, Sean M. and {K{\'o}sp{\'a}l}, {\'A}gnes and {Wilner}, David J. and {Chiang}, Eugene and {Armitage}, Philip J. and {Bai}, Xue-ning},
        title = "{A Three-dimensional View of Turbulence: Constraints on Turbulent Motions in the HD 163296 Protoplanetary Disk Using DCO$^{+}$}",
      journal = {\apj},
         year = 2017,
        month = jul,
       volume = {843},
       number = {2},
          eid = {150},
        pages = {150},
          doi = {10.3847/1538-4357/aa79f9},
archivePrefix = {arXiv},
       eprint = {1706.04504},
 primaryClass = {astro-ph.EP},
       adsurl = {https://ui.adsabs.harvard.edu/abs/2017ApJ...843..150F}
}

@ARTICLE{2012ApJ...744..118J,
       author = {{Juh{\'a}sz}, A. and {Dullemond}, C.~P. and {van Boekel}, R. and {Bouwman}, J. and {{\'A}brah{\'a}m}, P. and {Acosta-Pulido}, J.~A. and {Henning}, Th. and {K{\'o}sp{\'a}l}, A. and {Sicilia-Aguilar}, A. and {Jones}, A. and {Mo{\'o}r}, A. and {Mosoni}, L. and {Reg{\'a}ly}, Zs. and {Szokoly}, Gy. and {Sipos}, N.},
        title = "{The 2008 Outburst of EX Lup{\textemdash}Silicate Crystals in Motion}",
      journal = {\apj},
         year = 2012,
        month = jan,
       volume = {744},
       number = {2},
          eid = {118},
        pages = {118},
          doi = {10.1088/0004-637X/744/2/118},
archivePrefix = {arXiv},
       eprint = {1110.3754},
 primaryClass = {astro-ph.SR},
       adsurl = {https://ui.adsabs.harvard.edu/abs/2012ApJ...744..118J}
}

@ARTICLE{2011ApJ...733L..32P,
       author = {{Poteet}, Charles A. and {Megeath}, S. Thomas and {Watson}, Dan M. and {Calvet}, Nuria and {Remming}, Ian S. and {McClure}, Melissa K. and {Sargent}, Benjamin A. and {Fischer}, William J. and {Furlan}, Elise and {Allen}, Lori E. and {Bjorkman}, Jon E. and {Hartmann}, Lee and {Muzerolle}, James and {Tobin}, John J. and {Ali}, Babar},
        title = "{A Spitzer Infrared Spectrograph Detection of Crystalline Silicates in a Protostellar Envelope}",
      journal = {\apjl},
         year = 2011,
        month = jun,
       volume = {733},
       number = {2},
          eid = {L32},
        pages = {L32},
          doi = {10.1088/2041-8205/733/2/L32},
archivePrefix = {arXiv},
       eprint = {1104.4498},
 primaryClass = {astro-ph.SR},
       adsurl = {https://ui.adsabs.harvard.edu/abs/2011ApJ...733L..32P}
}

@ARTICLE{1997ApJ...479..740F,
       author = {{Falcke}, Heino and {Melia}, Fulvio},
        title = "{Accretion Disk Evolution with Wind Infall. I. General Solution and Application to Sagittarius A*}",
      journal = {\apj},
         year = 1997,
        month = apr,
       volume = {479},
       number = {2},
        pages = {740-751},
          doi = {10.1086/303893},
archivePrefix = {arXiv},
       eprint = {astro-ph/9611095},
 primaryClass = {astro-ph},
       adsurl = {https://ui.adsabs.harvard.edu/abs/1997ApJ...479..740F}
}

@ARTICLE{2019ApJ...886..103O,
       author = {{Ohashi}, Satoshi and {Kataoka}, Akimasa},
        title = "{Radial Variations in Grain Sizes and Dust Scale Heights in the Protoplanetary Disk around HD 163296 Revealed by ALMA Polarization Observations}",
      journal = {\apj},
         year = 2019,
        month = dec,
       volume = {886},
       number = {2},
          eid = {103},
        pages = {103},
          doi = {10.3847/1538-4357/ab5107},
archivePrefix = {arXiv},
       eprint = {1910.12868},
 primaryClass = {astro-ph.EP},
       adsurl = {https://ui.adsabs.harvard.edu/abs/2019ApJ...886..103O}
}

@INPROCEEDINGS{2017ASSL..445..161B,
       author = {{Bizzarro}, Martin and {Connelly}, James N. and {Krot}, Alexander N.},
        title = "{Chondrules: Ubiquitous Chondritic Solids Tracking the Evolution of the Solar Protoplanetary Disk}",
    booktitle = {Formation, Evolution, and Dynamics of Young Solar Systems},
         year = 2017,
       editor = {{Pessah}, Martin and {Gressel}, Oliver},
       series = {Astrophysics and Space Science Library},
       volume = {445},
        month = jan,
        pages = {161},
          doi = {10.1007/978-3-319-60609-5\_6},
       adsurl = {https://ui.adsabs.harvard.edu/abs/2017ASSL..445..161B}
}

@ARTICLE{2008A&A...480..859B,
       author = {{Brauer}, F. and {Dullemond}, C.~P. and {Henning}, Th.},
        title = "{Coagulation, fragmentation and radial motion of solid particles in protoplanetary disks}",
      journal = {\aap},
         year = 2008,
        month = mar,
       volume = {480},
       number = {3},
        pages = {859-877},
          doi = {10.1051/0004-6361:20077759},
archivePrefix = {arXiv},
       eprint = {0711.2192},
 primaryClass = {astro-ph},
       adsurl = {https://ui.adsabs.harvard.edu/abs/2008A&A...480..859B}
}

@ARTICLE{2018ApJ...864..133T,
       author = {{Teague}, Richard and {Henning}, Thomas and {Guilloteau}, St{\'e}phane and {Bergin}, Edwin A. and {Semenov}, Dmitry and {Dutrey}, Anne and {Flock}, Mario and {Gorti}, Uma and {Birnstiel}, Tilman},
        title = "{Temperature, Mass, and Turbulence: A Spatially Resolved Multiband Non-LTE Analysis of CS in TW Hya}",
      journal = {\apj},
         year = 2018,
        month = sep,
       volume = {864},
       number = {2},
          eid = {133},
        pages = {133},
          doi = {10.3847/1538-4357/aad80e},
archivePrefix = {arXiv},
       eprint = {1808.01768},
 primaryClass = {astro-ph.EP},
       adsurl = {https://ui.adsabs.harvard.edu/abs/2018ApJ...864..133T}
}

@ARTICLE{2007A&A...469.1169B,
       author = {{Brauer}, F. and {Dullemond}, C.~P. and {Johansen}, A. and {Henning}, Th. and {Klahr}, H. and {Natta}, A.},
        title = "{Survival of the mm-cm size grain population observed in protoplanetary disks}",
      journal = {\aap},
         year = 2007,
        month = jul,
       volume = {469},
       number = {3},
        pages = {1169-1182},
          doi = {10.1051/0004-6361:20066865},
archivePrefix = {arXiv},
       eprint = {0704.2332},
 primaryClass = {astro-ph},
       adsurl = {https://ui.adsabs.harvard.edu/abs/2007A&A...469.1169B}
}

@ARTICLE{2017A&A...602A..52W,
       author = {{Wijnen}, T.~P.~G. and {Pols}, O.~R. and {Pelupessy}, F.~I. and {Portegies Zwart}, S.},
        title = "{Characterising face-on accretion onto and the subsequent contraction of protoplanetary discs}",
      journal = {\aap},
         year = 2017,
        month = jun,
       volume = {602},
          eid = {A52},
        pages = {A52},
          doi = {10.1051/0004-6361/201630221},
archivePrefix = {arXiv},
       eprint = {1702.04383},
 primaryClass = {astro-ph.SR},
       adsurl = {https://ui.adsabs.harvard.edu/abs/2017A&A...602A..52W}
}

@ARTICLE{2019ApJ...872..112V,
       author = {{van der Marel}, Nienke and {Dong}, Ruobing and {di Francesco}, James and {Williams}, Jonathan P. and {Tobin}, John},
        title = "{Protoplanetary Disk Rings and Gaps across Ages and Luminosities}",
      journal = {\apj},
         year = 2019,
        month = feb,
       volume = {872},
       number = {1},
          eid = {112},
        pages = {112},
          doi = {10.3847/1538-4357/aafd31},
archivePrefix = {arXiv},
       eprint = {1901.03680},
 primaryClass = {astro-ph.EP},
       adsurl = {https://ui.adsabs.harvard.edu/abs/2019ApJ...872..112V}
}

@ARTICLE{2020MNRAS.493.4022L,
       author = {{Liffman}, Kurt and {Bryan}, Geoffrey and {Hutchison}, Mark and {Maddison}, Sarah T.},
        title = "{Infrared variability due to magnetic pressure-driven jets, dust ejection and quasi-puffed-up inner rims}",
      journal = {\mnras},
         year = 2020,
        month = apr,
       volume = {493},
       number = {3},
        pages = {4022-4038},
          doi = {10.1093/mnras/staa402},
archivePrefix = {arXiv},
       eprint = {2002.08432},
 primaryClass = {astro-ph.SR},
       adsurl = {https://ui.adsabs.harvard.edu/abs/2020MNRAS.493.4022L}
}

@ARTICLE{1981PThPS..70...35H,
       author = {{Hayashi}, C.},
        title = "{Structure of the Solar Nebula, Growth and Decay of Magnetic Fields and Effects of Magnetic and Turbulent Viscosities on the Nebula}",
      journal = {Progress of Theoretical Physics Supplement},
         year = 1981,
        month = jan,
       volume = {70},
        pages = {35-53},
          doi = {10.1143/PTPS.70.35},
       adsurl = {https://ui.adsabs.harvard.edu/abs/1981PThPS..70...35H}
}

@ARTICLE{1986Icar...67..375N,
       author = {{Nakagawa}, Y. and {Sekiya}, M. and {Hayashi}, C.},
        title = "{Settling and growth of dust particles in a laminar phase of a low-mass solar nebula}",
      journal = {\icarus},
         year = 1986,
        month = sep,
       volume = {67},
       number = {3},
        pages = {375-390},
          doi = {10.1016/0019-1035(86)90121-1},
       adsurl = {https://ui.adsabs.harvard.edu/abs/1986Icar...67..375N}
}

@ARTICLE{2020Natur.586..228S,
       author = {{Segura-Cox}, Dominique M. and {Schmiedeke}, Anika and {Pineda}, Jaime E. and {Stephens}, Ian W. and {Fern{\'a}ndez-L{\'o}pez}, Manuel and {Looney}, Leslie W. and {Caselli}, Paola and {Li}, Zhi-Yun and {Mundy}, Lee G. and {Kwon}, Woojin and {Harris}, Robert J.},
        title = "{Four annular structures in a protostellar disk less than 500,000 years old}",
      journal = {\nat},
         year = 2020,
        month = oct,
       volume = {586},
       number = {7828},
        pages = {228-231},
          doi = {10.1038/s41586-020-2779-6},
archivePrefix = {arXiv},
       eprint = {2010.03657},
 primaryClass = {astro-ph.EP},
       adsurl = {https://ui.adsabs.harvard.edu/abs/2020Natur.586..228S}
}

@ARTICLE{1977MNRAS.180...57W,
       author = {{Weidenschilling}, S.~J.},
        title = "{Aerodynamics of solid bodies in the solar nebula.}",
      journal = {\mnras},
         year = 1977,
        month = jul,
       volume = {180},
        pages = {57-70},
          doi = {10.1093/mnras/180.2.57},
       adsurl = {https://ui.adsabs.harvard.edu/abs/1977MNRAS.180...57W}
}

@ARTICLE{2021SciA....7.5967W,
       author = {{Weiss}, Benjamin P. and {Bai}, Xue-Ning and {Fu}, Roger R.},
        title = "{History of the solar nebula from meteorite paleomagnetism}",
      journal = {Science Advances},
         year = 2021,
        month = jan,
       volume = {7},
       number = {1},
        pages = {eaba5967},
          doi = {10.1126/sciadv.aba5967},
archivePrefix = {arXiv},
       eprint = {2103.02011},
 primaryClass = {astro-ph.EP},
       adsurl = {https://ui.adsabs.harvard.edu/abs/2021SciA....7.5967W}
}

@ARTICLE{2016ApJ...821...80B,
       author = {{Bai}, Xue-Ning},
        title = "{Towards a Global Evolutionary Model of Protoplanetary Disks}",
      journal = {\apj},
         year = 2016,
        month = apr,
       volume = {821},
       number = {2},
          eid = {80},
        pages = {80},
          doi = {10.3847/0004-637X/821/2/80},
archivePrefix = {arXiv},
       eprint = {1603.00484},
 primaryClass = {astro-ph.EP},
       adsurl = {https://ui.adsabs.harvard.edu/abs/2016ApJ...821...80B}
}

@ARTICLE{2021SciA....7.6928B,
       author = {{Borlina}, Cau{\^e} S. and {Weiss}, Benjamin P. and {Bryson}, James F.~J. and {Bai}, Xue-Ning and {Lima}, Eduardo A. and {Chatterjee}, Nilanjan and {Mansbach}, Elias N.},
        title = "{Paleomagnetic evidence for a disk substructure in the early solar system}",
      journal = {Science Advances},
         year = 2021,
        month = oct,
       volume = {7},
       number = {42},
          eid = {eabj6928},
        pages = {eabj6928},
          doi = {10.1126/sciadv.abj6928},
archivePrefix = {arXiv},
       eprint = {2110.09500},
 primaryClass = {astro-ph.EP},
       adsurl = {https://ui.adsabs.harvard.edu/abs/2021SciA....7.6928B}
}

@ARTICLE{2019ApJ...887..156A,
       author = {{{\'A}brah{\'a}m}, P. and {Chen}, L. and {K{\'o}sp{\'a}l}, {\'A}. and {Bouwman}, J. and {Carmona}, A. and {Haas}, M. and {Sicilia-Aguilar}, A. and {Sobrino Figaredo}, C. and {van Boekel}, R. and {Varga}, J.},
        title = "{Spectral Evolution and Radial Dust Transport in the Prototype Young Eruptive System EX Lup}",
      journal = {\apj},
         year = 2019,
        month = dec,
       volume = {887},
       number = {2},
          eid = {156},
        pages = {156},
          doi = {10.3847/1538-4357/ab521d},
archivePrefix = {arXiv},
       eprint = {1910.13180},
 primaryClass = {astro-ph.SR},
       adsurl = {https://ui.adsabs.harvard.edu/abs/2019ApJ...887..156A}
}

@ARTICLE{2023ApJ...945L...7K,
       author = {{K{\'o}sp{\'a}l}, {\'A}gnes and {{\'A}brah{\'a}m}, P{\'e}ter and {Diehl}, Lindsey and {Banzatti}, Andrea and {Bouwman}, Jeroen and {Chen}, Lei and {Cruz-S{\'a}enz de Miera}, Fernando and {Green}, Joel D. and {Henning}, Thomas and {Rab}, Christian},
        title = "{JWST/MIRI Spectroscopy of the Disk of the Young Eruptive Star EX Lup in Quiescence}",
      journal = {\apjl},
         year = 2023,
        month = mar,
       volume = {945},
       number = {1},
          eid = {L7},
        pages = {L7},
          doi = {10.3847/2041-8213/acb58a},
archivePrefix = {arXiv},
       eprint = {2301.08770},
 primaryClass = {astro-ph.SR},
       adsurl = {https://ui.adsabs.harvard.edu/abs/2023ApJ...945L...7K}
}

@ARTICLE{1977ApJ...217..693H,
       author = {{Herbig}, G.~H.},
        title = "{Eruptive phenomena in early stellar evolution.}",
      journal = {\apj},
         year = 1977,
        month = nov,
       volume = {217},
        pages = {693-715},
          doi = {10.1086/155615},
       adsurl = {https://ui.adsabs.harvard.edu/abs/1977ApJ...217..693H}
}

@ARTICLE{1992Icar..100..608L,
       author = {{Liffman}, Kurt},
        title = "{The formation of chondrules via ablation}",
      journal = {\icarus},
         year = 1992,
        month = dec,
       volume = {100},
       number = {2},
        pages = {608-620},
          doi = {10.1016/0019-1035(92)90121-M},
       adsurl = {https://ui.adsabs.harvard.edu/abs/1992Icar..100..608L}
}

@ARTICLE{1995Icar..116..275L,
       author = {{Liffman}, Kurt and {Brown}, Michael},
        title = "{The motion and size sorting of particles ejected from a protostellar accretion disk.}",
      journal = {\icarus},
         year = 1995,
        month = aug,
       volume = {116},
       number = {2},
        pages = {275-290},
          doi = {10.1006/icar.1995.1126},
       adsurl = {https://ui.adsabs.harvard.edu/abs/1995Icar..116..275L}
}

@ARTICLE{1996Sci...271.1545S,
       author = {{Shu}, Frank H. and {Shang}, Hsien and {Lee}, Typhoon},
        title = "{Toward an Astrophysical Theory of Chondrites}",
      journal = {Science},
         year = 1996,
        month = mar,
       volume = {271},
       number = {5255},
        pages = {1545-1552},
          doi = {10.1126/science.271.5255.1545},
       adsurl = {https://ui.adsabs.harvard.edu/abs/1996Sci...271.1545S}
}

@ARTICLE{2012ApJ...758..100B,
       author = {{Bans}, Alissa and {K{\"o}nigl}, Arieh},
        title = "{A Disk-wind Model for the Near-infrared Excess Emission in Protostars}",
      journal = {\apj},
         year = 2012,
        month = oct,
       volume = {758},
       number = {2},
          eid = {100},
        pages = {100},
          doi = {10.1088/0004-637X/758/2/100},
archivePrefix = {arXiv},
       eprint = {1207.1508},
 primaryClass = {astro-ph.SR},
       adsurl = {https://ui.adsabs.harvard.edu/abs/2012ApJ...758..100B}
}

@ARTICLE{2016ApJ...821....3M,
       author = {{Miyake}, Tomoya and {Suzuki}, Takeru K. and {Inutsuka}, Shu-ichiro},
        title = "{Dust Dynamics in Protoplanetary Disk Winds Driven by Magnetorotational Turbulence: A Mechanism for Floating Dust Grains with Characteristic Sizes}",
      journal = {\apj},
         year = 2016,
        month = apr,
       volume = {821},
       number = {1},
          eid = {3},
        pages = {3},
          doi = {10.3847/0004-637X/821/1/3},
archivePrefix = {arXiv},
       eprint = {1505.03704},
 primaryClass = {astro-ph.EP},
       adsurl = {https://ui.adsabs.harvard.edu/abs/2016ApJ...821....3M}
}

@ARTICLE{2019ApJ...882...33G,
       author = {{Giacalone}, Steven and {Teitler}, Seth and {K{\"o}nigl}, Arieh and {Krijt}, Sebastiaan and {Ciesla}, Fred J.},
        title = "{Dust Transport and Processing in Centrifugally Driven Protoplanetary Disk Winds}",
      journal = {\apj},
         year = 2019,
        month = sep,
       volume = {882},
       number = {1},
          eid = {33},
        pages = {33},
          doi = {10.3847/1538-4357/ab311a},
archivePrefix = {arXiv},
       eprint = {1907.04961},
 primaryClass = {astro-ph.SR},
       adsurl = {https://ui.adsabs.harvard.edu/abs/2019ApJ...882...33G}
}

@ARTICLE{2005ApJ...630.1088K,
       author = {{Krauss}, O. and {Wurm}, G.},
        title = "{Photophoresis and the Pile-up of Dust in Young Circumstellar Disks}",
      journal = {\apj},
         year = 2005,
        month = sep,
       volume = {630},
       number = {2},
        pages = {1088-1092},
          doi = {10.1086/432087},
       adsurl = {https://ui.adsabs.harvard.edu/abs/2005ApJ...630.1088K}
}

@ARTICLE{2022ApJ...940..117Z,
       author = {{Zhou}, Tingtao and {Deng}, Hong-Ping and {Chen}, Yi-Xian and {Lin}, Douglas N.~C.},
        title = "{Turbulent Transport of Dust Particles in Protostellar Disks: The Effect of Upstream Diffusion}",
      journal = {\apj},
         year = 2022,
        month = dec,
       volume = {940},
       number = {2},
          eid = {117},
        pages = {117},
          doi = {10.3847/1538-4357/ac9bf6},
archivePrefix = {arXiv},
       eprint = {2210.10815},
 primaryClass = {astro-ph.EP},
       adsurl = {https://ui.adsabs.harvard.edu/abs/2022ApJ...940..117Z}
}

@INPROCEEDINGS{1996cppd.proc..285L,
       author = {{Liffman}, K. and {Brown}, M.~J.~I.},
        title = "{The Protostellar Jet Model of Chondrule Formation}",
    booktitle = {Chondrules and the Protoplanetary Disk},
         year = 1996,
        month = jan,
        pages = {285-302},
          doi = {10.48550/arXiv.astro-ph/0602383},
archivePrefix = {arXiv},
       eprint = {astro-ph/0602383},
 primaryClass = {astro-ph},
       adsurl = {https://ui.adsabs.harvard.edu/abs/1996cppd.proc..285L}
}

@ARTICLE{2021MNRAS.500..506V,
       author = {{Vinkovi{\'c}}, Dejan and {{\v{C}}emelji{\'c}}, Miljenko},
        title = "{Inner dusty regions of protoplanetary discs - II. Dust dynamics driven by radiation pressure and disc winds}",
      journal = {\mnras},
         year = 2021,
        month = jan,
       volume = {500},
       number = {1},
        pages = {506-519},
          doi = {10.1093/mnras/staa3272},
archivePrefix = {arXiv},
       eprint = {2010.09384},
 primaryClass = {astro-ph.EP},
       adsurl = {https://ui.adsabs.harvard.edu/abs/2021MNRAS.500..506V}
}

@ARTICLE{2016MNRAS.458.2140C,
       author = {{Cuello}, N. and {Gonzalez}, J. -F. and {Pignatale}, F.~C.},
        title = "{Effects of photophoresis on the dust distribution in a 3D protoplanetary disc}",
      journal = {\mnras},
         year = 2016,
        month = may,
       volume = {458},
       number = {2},
        pages = {2140-2149},
          doi = {10.1093/mnras/stw396},
archivePrefix = {arXiv},
       eprint = {1601.03662},
 primaryClass = {astro-ph.EP},
       adsurl = {https://ui.adsabs.harvard.edu/abs/2016MNRAS.458.2140C}
}

@ARTICLE{2012A&A...538A.114P,
       author = {{Pinilla}, P. and {Birnstiel}, T. and {Ricci}, L. and {Dullemond}, C.~P. and {Uribe}, A.~L. and {Testi}, L. and {Natta}, A.},
        title = "{Trapping dust particles in the outer regions of protoplanetary disks}",
      journal = {\aap},
         year = 2012,
        month = feb,
       volume = {538},
          eid = {A114},
        pages = {A114},
          doi = {10.1051/0004-6361/201118204},
archivePrefix = {arXiv},
       eprint = {1112.2349},
 primaryClass = {astro-ph.EP},
       adsurl = {https://ui.adsabs.harvard.edu/abs/2012A&A...538A.114P}
}

@ARTICLE{2016MNRAS.462.1137L,
       author = {{Liffman}, Kurt and {Cuello}, Nicolas and {Paterson}, David A.},
        title = "{A unified framework for producing CAI melting, Wark-Lovering rims and bowl-shaped CAIs}",
      journal = {\mnras},
         year = 2016,
        month = oct,
       volume = {462},
       number = {2},
        pages = {1137-1163},
          doi = {10.1093/mnras/stw1563},
archivePrefix = {arXiv},
       eprint = {1606.07539},
 primaryClass = {astro-ph.EP},
       adsurl = {https://ui.adsabs.harvard.edu/abs/2016MNRAS.462.1137L}
}

@ARTICLE{2014MNRAS.440.3545H,
       author = {{Hansen}, Bradley M.~S.},
        title = "{The circulation of dust in protoplanetary discs and the initial conditions of planet formation}",
      journal = {\mnras},
         year = 2014,
        month = jun,
       volume = {440},
       number = {4},
        pages = {3545-3556},
          doi = {10.1093/mnras/stu471},
archivePrefix = {arXiv},
       eprint = {1403.6552},
 primaryClass = {astro-ph.EP},
       adsurl = {https://ui.adsabs.harvard.edu/abs/2014MNRAS.440.3545H}
}

@ARTICLE{2020E&PSL.53516088L,
       author = {{Larsen}, K.~K. and {Wielandt}, D. and {Schiller}, M. and {Krot}, A.~N. and {Bizzarro}, M.},
        title = "{Episodic formation of refractory inclusions in the Solar System and their presolar heritage}",
      journal = {Earth and Planetary Science Letters},
         year = 2020,
        month = apr,
       volume = {535},
          eid = {116088},
        pages = {116088},
          doi = {10.1016/j.epsl.2020.116088},
       adsurl = {https://ui.adsabs.harvard.edu/abs/2020E&PSL.53516088L}
}

@ARTICLE{2012M&PS...47.1922S,
       author = {{Salmeron}, Raquel and {Ireland}, Trevor},
        title = "{The role of protostellar jets in star formation and the evolution of the early solar system: Astrophysical and meteoritical perspectives}",
      journal = {M\&PS},
         year = 2012,
        month = dec,
       volume = {47},
       number = {12},
        pages = {1922-1940},
          doi = {10.1111/maps.12029},
       adsurl = {https://ui.adsabs.harvard.edu/abs/2012M&PS...47.1922S}
}

@ARTICLE{2011Sci...332.1528M,
       author = {{McKeegan}, K.~D. and {Kallio}, A.~P.~A. and {Heber}, V.~S. and {Jarzebinski}, G. and {Mao}, P.~H. and {Coath}, C.~D. and {Kunihiro}, T. and {Wiens}, R.~C. and {Nordholt}, J.~E. and {Moses}, R.~W. and {Reisenfeld}, D.~B. and {Jurewicz}, A.~J.~G. and {Burnett}, D.~S.},
        title = "{The Oxygen Isotopic Composition of the Sun Inferred from Captured Solar Wind}",
      journal = {Science},
         year = 2011,
        month = jun,
       volume = {332},
       number = {6037},
        pages = {1528},
          doi = {10.1126/science.1204636},
       adsurl = {https://ui.adsabs.harvard.edu/abs/2011Sci...332.1528M}
}

@ARTICLE{2010ApJ...725.1421H,
       author = {{Hu}, Renyu},
        title = "{Transport of the First Rocks of the Solar System by X-winds}",
      journal = {\apj},
         year = 2010,
        month = dec,
       volume = {725},
       number = {2},
        pages = {1421-1428},
          doi = {10.1088/0004-637X/725/2/1421},
archivePrefix = {arXiv},
       eprint = {1010.2532},
 primaryClass = {astro-ph.EP},
       adsurl = {https://ui.adsabs.harvard.edu/abs/2010ApJ...725.1421H}
}

@ARTICLE{2022ApJ...928...92K,
       author = {{Kuznetsova}, Aleksandra and {Bae}, Jaehan and {Hartmann}, Lee and {Mac Low}, Mordecai-Mark},
        title = "{Anisotropic Infall and Substructure Formation in Embedded Disks}",
      journal = {\apj},
         year = 2022,
        month = mar,
       volume = {928},
       number = {1},
          eid = {92},
        pages = {92},
          doi = {10.3847/1538-4357/ac54a8},
archivePrefix = {arXiv},
       eprint = {2202.05301},
 primaryClass = {astro-ph.EP},
       adsurl = {https://ui.adsabs.harvard.edu/abs/2022ApJ...928...92K}
}

@ARTICLE{2000Icar..143..106L,
       author = {{Liffman}, Kurt and {Toscano}, Maurizio},
        title = "{Chondrule Fine-Grained Mantle Formation by Hypervelocity Impact of Chondrules with a Dusty Gas}",
      journal = {\icarus},
         year = 2000,
        month = jan,
       volume = {143},
       number = {1},
        pages = {106-125},
          doi = {10.1006/icar.1999.6249},
       adsurl = {https://ui.adsabs.harvard.edu/abs/2000Icar..143..106L}
}

@ARTICLE{2016MNRAS.461..742H,
       author = {{Hutchison}, Mark A. and {Price}, Daniel J. and {Laibe}, Guillaume and {Maddison}, Sarah T.},
        title = "{On dust entrainment in photoevaporative winds}",
      journal = {\mnras},
         year = 2016,
        month = sep,
       volume = {461},
       number = {1},
        pages = {742-759},
          doi = {10.1093/mnras/stw1126},
archivePrefix = {arXiv},
       eprint = {1603.02899},
 primaryClass = {astro-ph.SR},
       adsurl = {https://ui.adsabs.harvard.edu/abs/2016MNRAS.461..742H}
}

@ARTICLE{2023ApJ...946L..34H,
       author = {{Hellmann}, Jan L. and {Schneider}, Jonas M. and {W{\"o}lfer}, Elias and {Dr{\k{a}}{\.z}kowska}, Joanna and {Jansen}, Christian A. and {Hopp}, Timo and {Burkhardt}, Christoph and {Kleine}, Thorsten},
        title = "{Origin of Isotopic Diversity among Carbonaceous Chondrites}",
      journal = {\apjl},
         year = 2023,
        month = apr,
       volume = {946},
       number = {2},
          eid = {L34},
        pages = {L34},
          doi = {10.3847/2041-8213/acc102},
archivePrefix = {arXiv},
       eprint = {2303.04173},
 primaryClass = {astro-ph.EP},
       adsurl = {https://ui.adsabs.harvard.edu/abs/2023ApJ...946L..34H}
}

@INPROCEEDINGS{1997ASPC..121..241L,
       author = {{Li}, J. and {Wickramasinghe}, D.~T.},
        title = "{Disc Accretion Onto Magnetic Stars: Slow Rotator and Propeller}",
    booktitle = {IAU Colloq. 163: Accretion Phenomena and Related Outflows},
         year = 1997,
       editor = {{Wickramasinghe}, D.~T. and {Bicknell}, G.~V. and {Ferrario}, L.},
       series = {Astronomical Society of the Pacific Conference Series},
       volume = {121},
        month = jan,
        pages = {241},
       adsurl = {https://ui.adsabs.harvard.edu/abs/1997ASPC..121..241L}
}

\appendix

\section{Mass and Angular Conservation Equations}
\label{sec:infall_eqns}
\subsection{Mass Conservation}
\label{subsec:mass_con}
\begin{figure}[H]
\centering
\includegraphics[width=\textwidth]{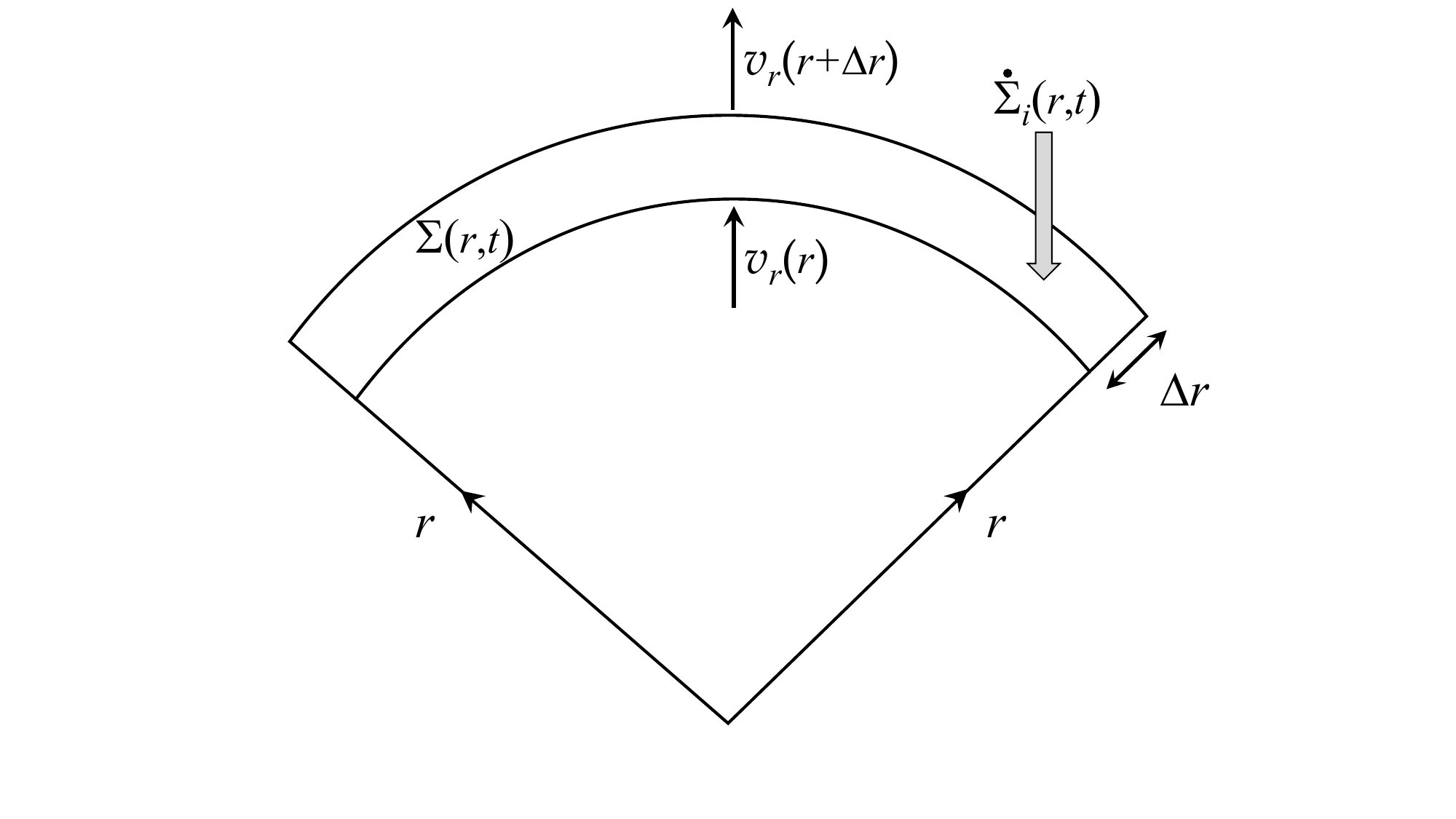}
\caption{Schematic of a disc differential ring that illustrates the disc mass conservation equation with infall.}
\label{fig:mass_cons}
\end{figure}
From Figure (\ref{fig:mass_cons}),
\begin{equation}
\begin{aligned}
2 \pi r \Delta r \frac{\partial \Sigma}{\partial t} &= 2 \pi r v_{\rm r}(r,t) \Sigma(r,t)  \\ &- 2 \pi (r + \Delta r) v_{\rm r}(r+\Delta r,t) \Sigma(r + \Delta r, t) \\ &+ 2 \pi r \Delta r \dot{\Sigma}_{\rm i} (r,t)
\end{aligned}
\end{equation}
dividing both sides by $2 \pi r \Delta r$ and allowing $\Delta r \rightarrow 0$ gives the disc mass conservation:
\begin{equation}
    \frac{\partial \Sigma}{\partial t} = -\frac{1}{r}\frac{\partial \left(r v_{\rm r} \Sigma \right)}{\partial r} + \dot{\Sigma}_{\rm i} \ .
    \label{eq:mass_con_2}
\end{equation}

\subsection{Angular Momentum Conservation}
\label{subsec:ang_mom_con}

\begin{figure}[H]
\centering
\includegraphics[width=\textwidth]{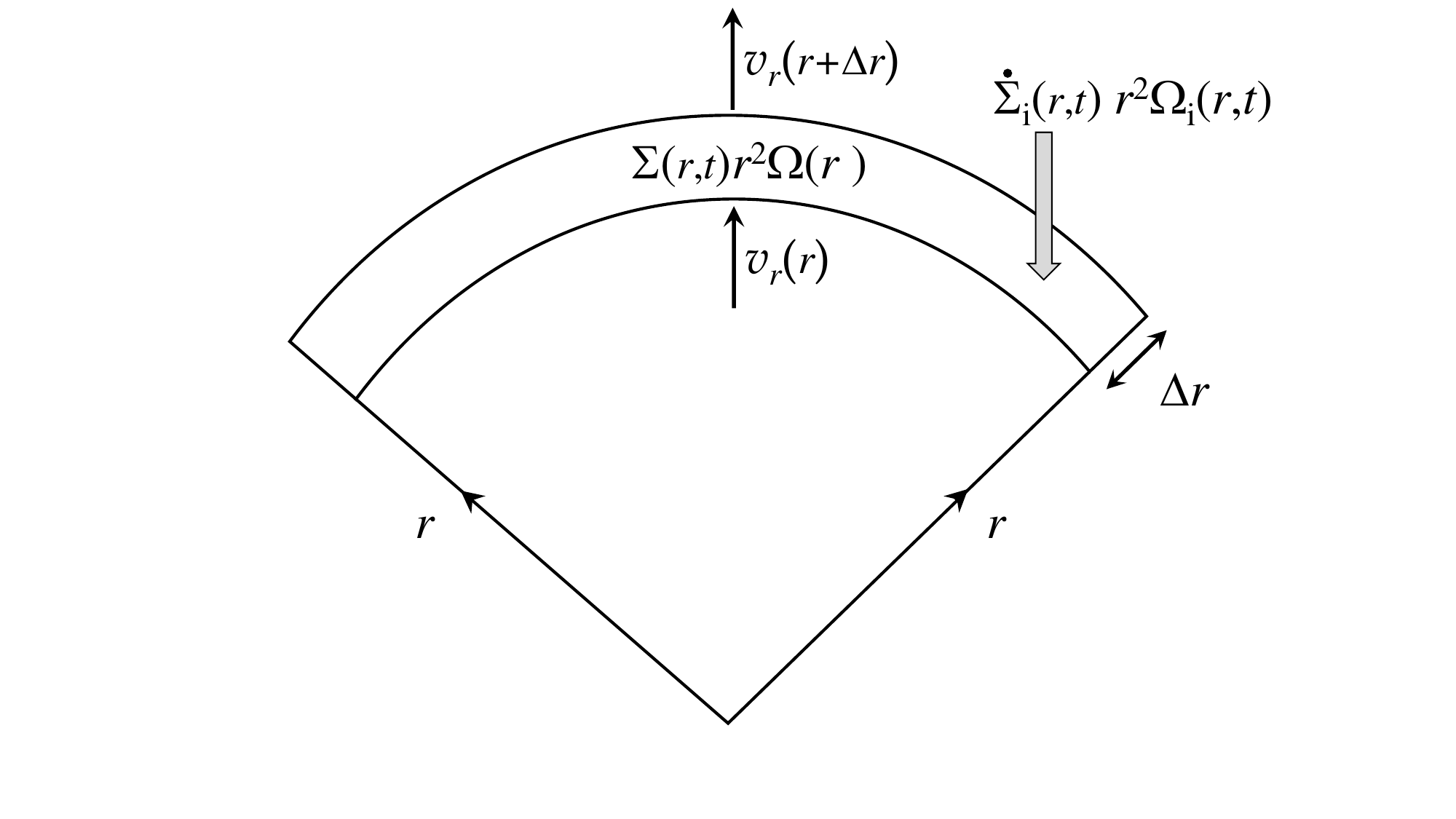}
\caption{Schematic of a disc differential ring that illustrates the disc angular momentum conservation equation with infall.}
\label{fig:ang_mom_cons}
\end{figure}

From Figure (\ref{fig:ang_mom_cons}),
\begin{equation}
\begin{aligned}
&2 \pi r \Delta r \frac{\partial \left( \Sigma r^2 \Omega \right)}{\partial t} 
= 2 \pi r v_{\rm r}(r,t) \Sigma(r,t) r^2 \Omega   \\ 
&- 2 \pi (r + \Delta r) v_{\rm r}(r+\Delta r,t) \Sigma(r + \Delta r, t) (r + \Delta r)^2 \Omega(r + \Delta r)  \\ 
&+ \Delta r \frac{\partial Q}{\partial r} 
+ 2 \pi r \Delta r \dot{\Sigma}_{\rm i} (r,t) r^2 \Omega_{\rm i}(r,t) \ ,
\end{aligned}
\end{equation}
which gives the angular momentum conservation equation:

\begin{equation}
\begin{aligned}
    \frac{\partial \left( \Sigma r^2 \Omega \right)}{\partial t} &= -\frac{1}{r}\frac{\partial \left(r v_{\rm r} \Sigma r^2 \Omega \right)}{\partial r} \\
    &+ \frac{1}{2 \pi r}\frac{\partial Q}{\partial r}+ \dot{\Sigma}_{\rm i} r^2 \Omega_{\rm i} \ .
    \label{eq:ang_mom_cons_2}
    \end{aligned}
\end{equation}

\section{Disc Accretion Speed}
\label{sec:disc_infall}
From equation (\ref{eq:mass_con_2}), 
\begin{equation}
    \frac{\partial(v_{\rm r} \Sigma r^3 \Omega)}{\partial r} = r^2 \Omega \left( r \dot{\Sigma}_{\rm i} - r \frac{\partial \Sigma}{\partial t}\right) + r v_{\rm r} \Sigma \frac{\partial (r^2 \Omega)}{\partial r} \ ,
\end{equation}
while
\begin{equation}
\begin{aligned}
\frac{\partial \left( \Sigma r^2 \Omega \right)}{\partial t} &= 
\frac{\partial \Sigma}{\partial t} r^2 \Omega \ .
\end{aligned}
\end{equation}

Combining these equations with equation (\ref{eq:ang_mom_cons_2})
\begin{equation}
\begin{aligned}
    r\frac{\partial \left( \Sigma r^2 \Omega \right)}{\partial t} &+ \frac{\partial \left(r v_{\rm r} \Sigma r^2 \Omega \right)}{\partial r}   \\
    &= \dot{\Sigma}_{\rm i} r^3 \Omega + r v_{\rm r}\Sigma\frac{\partial \left(  r^2 \Omega \right)}{\partial r} \\
    &=\frac{1}{2 \pi}\frac{\partial Q}{\partial r}+ \dot{\Sigma}_{\rm i} r^3 \Omega_{\rm i} \ ,
    \end{aligned}
\end{equation}

and so,
\begin{equation}
     r v_{\rm r}\Sigma\frac{\partial \left(  r^2 \Omega \right)}{\partial r} =
    \frac{1}{2 \pi}\frac{\partial Q}{\partial r}+ \dot{\Sigma}_{\rm i} r^3 \left( \Omega_{\rm i} - \Omega\right)\ ,
\end{equation}
or
\begin{equation}
     v_{\rm r}  = \frac{1}{r\Sigma \frac{\partial \left(  r^2 \Omega \right)}{\partial r}}\left(
    \frac{1}{2 \pi}\frac{\partial Q}{\partial r}+ \dot{\Sigma}_{\rm i} r^3 \left( \Omega_{\rm i} - \Omega\right)\right)\ .
\end{equation}
We apply our inviscid disc assumption: $Q\approx 0$ and obtain
\begin{equation}
\begin{aligned}
     v_{\rm r}  &= \frac{1}{r\Sigma \frac{\partial \left(  r^2 \Omega \right)}{\partial r}}\left(
     \dot{\Sigma}_{\rm i} r^3 \left( \Omega_{\rm i} - \Omega\right)\right) \\
     &= \frac{1}{ \frac{\partial \left(  r^2 \Omega \right)}{\partial r}}
     \frac{\dot{\Sigma}_{\rm i}}{\Sigma} r^2 \left( \Omega_{\rm i} - \Omega\right)\ .
    \end{aligned}     
\end{equation}

\section{Analytic Solutions of the Disc Equations for the Case $\Omega_{\rm i} \approx 0$ and Constant $\dot{\Sigma}_{\rm i}$ }
\label{sec:analytic solution}

In this case,  (\ref{eq:rad_drift2}) becomes
\begin{equation}
    v_{\rm r} = \frac{dr}{dt} = -2r\frac{\dot{\Sigma}_{\rm i}}{\Sigma} \ ,
    \label{eq:rad_drift3}
\end{equation}
while equation (\ref{eq:dSigmadt}) has the form
\begin{equation}
      \frac{\partial \Sigma}{\partial t} = 5 \dot{\Sigma}_{\rm i} \ .
      \label{eq:dSigmadt2}
\end{equation}
We can eliminate $\dot{\Sigma}_{\rm i}$ from both equations and obtain
\begin{equation}
      \frac{1}{\Sigma}\frac{\partial \Sigma}{\partial t} = \frac{-5}{2r}\frac{dr}{dt}  \ ,
      \label{eq:dSigmadt4}
\end{equation}
which has the solution
\begin{equation}
      \Sigma(r(t),t)  = \Sigma(r(t_0),t_0)\left( \frac{r(t_0)}{r(t)} \right)^\frac{5}{2} \ ,
      \label{eq:Sigmasolution}
\end{equation}
Substituting equation (\ref{eq:Sigmasolution}) into equation (\ref{eq:rad_drift3}) and using the assumption that $\dot{\Sigma}_{\rm i}$ is a constant gives
\begin{equation}
    r(t) = \frac{r(t_0)}{\left( 1 + 5\frac{\dot{\Sigma}_{\rm i}}{\Sigma(r(t_0),t_0)}(t-t_0) \right)^{2/5}} \ .
    \label{eq:r(t)2}
\end{equation}

Substituting equation (\ref{eq:r(t)2}) into equation (\ref{eq:Sigmasolution}) implies
\begin{equation}
    \Sigma(r(t),t) = \Sigma(r(t_0), t_0) + 5\dot{\Sigma}_{\rm i}(t-t_0) \ .
    \label{eq:Sigma(t)2}
\end{equation}

\section{Drift Equations}
\label{sec:drift_eqns}

From \citet{1986Icar...67..375N}, the radial drift speed for particles, $v_{\rm dr}$, is given by
\begin{equation}
v_{\rm dr} = \frac{\rho_{\rm g} \tau_{\rm s} \frac{{\rm d}p}{{\rm d}r}}
{\left(\rho_{\rm g}+\rho_{\rm d} \right)^2+\rho_{\rm g}^2\tau_{\rm s}^2\Omega^2} \ ,
\label{eq:vdr}
\end{equation}
with $\rho_{\rm g}$ the spatial gas mass density, $\rho_{\rm d}$ the spatial particle mass density, $\frac{{\rm d}p}{{\rm d}r}$ the gas radial pressure gradient, $\tau_{\rm s}$ the particle stopping time and $\Omega$ the Keplerian angular frequency.

Assuming Epstein drag, the stopping time is
\begin{equation}
\tau_{\rm s} \approx \frac{a_{\rm p} \rho_{\rm p}}{\left(1+\pi/8\right)\rho_{\rm g}v_{\rm th}} \ ,
\label{eq:tau_s}
\end{equation}
here, $a_{\rm p}$ is the particle radius, $\rho_{\rm p}$ is the particle density (set to  $ 3,000 \ {\rm kg} {\rm m}^{-3}$ ), the $1+\pi/8$ is a correction factor for the Epstein drag coefficient \citep{2000Icar..143..106L}, and $v_{\rm th}$ is the thermal gas speed, where
\begin{equation}
v_{\rm th} = \sqrt{\frac{8 {\rm k}_{\rm BP} T_{\rm g}}{\pi \mu {\rm m}_{\rm H}}} \ ,
\label{eq:vth}
\end{equation}
with ${\rm k}_{\rm BP}$ the Boltzmann constant (where the P notes Planck as the person who actually proposed the constant), $\mu$ the mean molecular mass ($\approx 2.3$), ${\rm m}_{\rm H}$ the hydrogen atom mass and $T_{\rm g}$ the temperature of the gas.

For our disc model, we assume that
\begin{equation}
    \rho_{\rm g}(r) = \rho_{\rm g0}\left( \frac{r_0}{r}\right)^\alpha \ ,
    \label{eq:rho_g}
\end{equation}
with $r_0 = 1 $ au, $\rho_{\rm g0}$ is set to $ 10^{-6} \ {\rm kg} \ {\rm m}^{-3}$ and $\alpha = 3$.
\begin{equation}
    T_{\rm g}(r) = T_{\rm g0}\left( \frac{r_0}{r}\right)^\beta \ ,
    \label{eq:T_g}
\end{equation}
with $T_{\rm g0} = 500$ K and $\beta = 1/2$

The values of $\rho_{\rm g0}, T_{\rm g0}, \alpha$ and $\beta$ are chosen to produce a disc surface density that is approximately similar to the Hayashi minimum mass surface density for the solar system protoplanetary disc \citep{1981PThPS..70...35H}:
\begin{equation}
    \Sigma_{\rm Hayashi}(r) = 1700\left( \frac{r_0}{r}\right)^{3/2} \ {\rm g} \ {\rm cm}^{-3} .
\end{equation}
In our case, we used the model surface density:
\begin{equation}
    \Sigma_{\rm model}(r) \approx 2 \rho_{\rm g}(r)h(r) \ ,
\end{equation}
where $h(r)$ is the scale height of the disc:
\begin{equation}
    h(r) \approx \sqrt{\frac{2 r^3 {\rm k}_{\rm BP} T_{\rm g}(r)}{ \mu {\rm m}_{\rm H} {\rm G} M_*}} \ .
\end{equation}
To determine the pressure gradient in equation (\ref{eq:vdr}), we note that the gas pressure is
\begin{equation}
    p(r) = \frac{ {\rm k}_{\rm BP} \rho_{\rm g}(r) T_{\rm g}(r)}{ \mu {\rm m}_{\rm H}} \ .
\end{equation}

Substituting equations (\ref{eq:rho_g}) and (\ref{eq:T_g}) gives 
\begin{equation}
    \frac{{\rm d}p}{{\rm d}r} = -\left( \alpha + \beta \right) \frac{ {\rm k}_{\rm BP} \rho_{\rm g}(r) T_{\rm g}(r)}{ \mu {\rm m}_{\rm H}r} \ .
    \label{eq:dpdr}
\end{equation}

Substituting this equation into equation (\ref{eq:vdr}), and ignoring the $\rho_{\rm d}/\rho_{\rm g}$ term as it is $\sim$ 0.01, gives
\begin{equation}
v_{\rm dr} \approx \frac{ -\left( \alpha + \beta \right) \tau_{\rm s} {\rm k}_{\rm BP} T_{\rm g}(r) } {\mu {\rm m}_{\rm H}r\left(1+\tau_{\rm s}^2\Omega^2\right)} \ ,
\label{eq:vdrII}
\end{equation}

From equations (\ref{eq:vth}) and (\ref{eq:T_g})
\begin{equation}
v_{\rm th} = \sqrt{\frac{8 {\rm k}_{\rm BP} T_{\rm g0}}{\pi \mu {\rm m}_{\rm H}}}\left( \frac{r_0}{r}\right)^{\beta/2} \equiv v_{\rm th0}\left( \frac{r_0}{r}\right)^{\beta/2} \ .
\label{eq:vth0}
\end{equation}
Similarly,
\begin{equation}
\tau_{\rm s} \approx \frac{a_{\rm p} \rho_{\rm p}}{\left(1+\pi/8\right)\rho_{\rm g0}v_{\rm th0}}\left(\frac{r_0}{r} \right)^{-\alpha - \beta/2} 
\equiv \tau_{\rm s0}\left(\frac{r_0}{r} \right)^{-\alpha - \beta/2} \ .
\label{eq:tau_s2}
\end{equation}
Combining equations (\ref{eq:T_g}), (\ref{eq:tau_s2}), and (\ref{eq:vdrII}) plus defining $\Omega_0^2 = {\rm G} M_*/r_0^3 $, gives
\begin{equation}
v_{\rm dr} \approx \frac{- \left( \alpha + \beta \right) \tau_{\rm s0} {\rm k}_{\rm BP} T_{\rm g0}\left(\frac{r_0}{r} \right)^{-\alpha + \beta/2 +1} } {\mu {\rm m}_{\rm H}r_0\left(1+\tau_{\rm s0}^2\Omega_0^2\left(\frac{r_0}{r} \right)^{-2\alpha - \beta +3}\right)} \ .
\label{eq:vdrIII}
\end{equation}

Defining 
\begin{equation}
    A = \frac{ \left( \alpha + \beta \right) \tau_{\rm s0} {\rm k}_{\rm BP} T_{\rm g0} } {\mu {\rm m}_{\rm H}r_0} \ ,
\end{equation}
and
\begin{equation}
    B = \tau_{\rm s0}^2\Omega_0^2 \ ,
\end{equation}
implies
\begin{equation}
v_{\rm dr} =  \frac{{\rm d}r}{{\rm d}t} \approx \frac{-A\left(\frac{r_0}{r} \right)^{-\alpha + \beta/2 +1} } {\left(1+B\left(\frac{r_0}{r} \right)^{-2\alpha - \beta +3}\right)} \ .
\label{eq:vdrIV}
\end{equation}
$\implies$
\begin{equation}
\left(\left(\frac{r_0}{r} \right)^{\alpha - \beta/2 -1}+B\left(\frac{r_0}{r} \right)^{-\alpha - 3\beta/2 +2}\right) \frac{{\rm d}r}{{\rm d}t} \approx -A  \ .
\end{equation}
$\implies$
\begin{equation}
\begin{aligned}
   & \left( \frac{1}{r_0}\right)^{-\alpha + \beta/2 +1} \frac{\left(r^{-\alpha + \beta/2 +2} -r_{t0}^{-\alpha + \beta/2 +2} \right)}{-\alpha + \beta/2 +2} \\
   &+B\left(\frac{1}{r_0} \right)^{\alpha + 3\beta/2 -2} 
   \frac{\left(r^{\alpha + 3\beta/2 -1} -r_{t0}^{\alpha + 3\beta/2 -1} \right)}{\alpha + 3\beta/2 -1}\\
   &\approx -A (t - t_0) \ ,
\end{aligned}
\label{eq:vdrV}
\end{equation}
where $t_0$ is the initial time and $r_{t0}$ is the initial position of the particle.

For $\alpha = 3$ and $\beta = 1/2$
\begin{equation}
\begin{aligned}
   &  \left(\frac{r_0}{r}\right)^{3/4} - \left(\frac{r_0}{r_{\rm t0}}\right)^{3/4} -
   \frac{3B}{11} \left( \left(\frac{r}{r_0}\right)^{11/4} - \left(\frac{r_{\rm t0}}{r_0}\right)^{11/4}\right)\\
   &\approx \frac{3A}{4r_0} (t - t_0) \ .
\end{aligned}
\label{eq:vdrVI}
\end{equation}
For particle radii less than approximately 1 m, $B \ll 1$, so
\begin{equation}
   \left(\frac{r_0}{r}\right)^{3/4} - \left(\frac{r_0}{r_{\rm t0}}\right)^{3/4} 
   \approx \frac{3A}{4r_0} (t - t_0) \ ,
\label{eq:vdrVII}
\end{equation}
and
\begin{equation}
t - t_0 \approx \tau_{\rm drift}
   \left(\left(\frac{r_0}{r}\right)^{3/4} - \left(\frac{r_0}{r_{\rm t0}}\right)^{3/4} \right)  \ ,
\end{equation}
with
\begin{equation}
\begin{aligned}
\tau_{\rm drift} &= \frac{4r_0}{3A} =
\frac{8 r^2_0 \mu m_{\rm H}}{ 21 \tau_{\rm s0} {\rm k}_{\rm BP} T_{\rm g0}} \\
&\approx 1.5\times10^{5}  \left( \frac{1 \ {\rm mm}}{a_{\rm p}} \right) {\rm yrs} \ .
\label{eq:tau_driftII}
\end{aligned}
\end{equation}

\end{document}